%% file: main.tex
\documentclass[letterpaper,twocolumn,10pt]{article}
\usepackage{usenix}

\usepackage{tikz}
\usepackage{amsmath}

\usepackage{booktabs}
\usepackage{adjustbox}
\usepackage{float}
\usepackage{pifont}
\usepackage{balance}
\usepackage{listings} 
\usepackage{color} 
\usepackage{xcolor}
\usepackage[draft]{minted}
\usepackage{transparent}
\usepackage{fancyvrb}
\usepackage{xurl}

\newcommand*\numcircledmod[1]{\raisebox{.5pt}{\textcircled{\raisebox{-.9pt} {#1}}}}
\newcommand{\cmark}{\ding{51}}
\newcommand{\xmark}{\ding{55}}

\makeatletter
\def\PYG@reset{\let\PYG@it=\relax \let\PYG@bf=\relax%
    \let\PYG@ul=\relax \let\PYG@tc=\relax%
    \let\PYG@bc=\relax \let\PYG@ff=\relax}
\def\PYG@tok#1{\csname PYG@tok@#1\endcsname}
\def\PYG@toks#1+{\ifx\relax#1\empty\else%
    \PYG@tok{#1}\expandafter\PYG@toks\fi}
\def\PYG@do#1{\PYG@bc{\PYG@tc{\PYG@ul{%
    \PYG@it{\PYG@bf{\PYG@ff{#1}}}}}}}
\def\PYG#1#2{\PYG@reset\PYG@toks#1+\relax+\PYG@do{#2}}

\@namedef{PYG@tok@w}{\def\PYG@tc##1{\textcolor[rgb]{0.73,0.73,0.73}{##1}}}
\@namedef{PYG@tok@c}{\def\PYG@tc##1{\textcolor[rgb]{0.13,0.55,0.13}{##1}}}
\@namedef{PYG@tok@cp}{\def\PYG@tc##1{\textcolor[rgb]{0.12,0.53,0.61}{##1}}}
\@namedef{PYG@tok@cs}{\let\PYG@bf=\textbf\def\PYG@tc##1{\textcolor[rgb]{0.55,0.00,0.55}{##1}}}
\@namedef{PYG@tok@s}{\def\PYG@tc##1{\textcolor[rgb]{0.80,0.33,0.33}{##1}}}
\@namedef{PYG@tok@sh}{\let\PYG@it=\textit\def\PYG@tc##1{\textcolor[rgb]{0.11,0.49,0.44}{##1}}}
\@namedef{PYG@tok@sr}{\def\PYG@tc##1{\textcolor[rgb]{0.11,0.49,0.44}{##1}}}
\@namedef{PYG@tok@sx}{\def\PYG@tc##1{\textcolor[rgb]{0.80,0.42,0.13}{##1}}}
\@namedef{PYG@tok@m}{\def\PYG@tc##1{\textcolor[rgb]{0.71,0.32,0.80}{##1}}}
\@namedef{PYG@tok@ow}{\def\PYG@tc##1{\textcolor[rgb]{0.55,0.00,0.55}{##1}}}
\@namedef{PYG@tok@k}{\let\PYG@bf=\textbf\def\PYG@tc##1{\textcolor[rgb]{0.55,0.00,0.55}{##1}}}
\@namedef{PYG@tok@kt}{\let\PYG@bf=\textbf\def\PYG@tc##1{\textcolor[rgb]{0.00,0.41,0.55}{##1}}}
\@namedef{PYG@tok@nc}{\let\PYG@bf=\textbf\def\PYG@tc##1{\textcolor[rgb]{0.00,0.55,0.27}{##1}}}
\@namedef{PYG@tok@ne}{\let\PYG@bf=\textbf\def\PYG@tc##1{\textcolor[rgb]{0.00,0.55,0.27}{##1}}}
\@namedef{PYG@tok@nf}{\def\PYG@tc##1{\textcolor[rgb]{0.00,0.55,0.27}{##1}}}
\@namedef{PYG@tok@nn}{\let\PYG@ul=\underline\def\PYG@tc##1{\textcolor[rgb]{0.00,0.55,0.27}{##1}}}
\@namedef{PYG@tok@nv}{\def\PYG@tc##1{\textcolor[rgb]{0.00,0.41,0.55}{##1}}}
\@namedef{PYG@tok@no}{\def\PYG@tc##1{\textcolor[rgb]{0.00,0.41,0.55}{##1}}}
\@namedef{PYG@tok@nd}{\def\PYG@tc##1{\textcolor[rgb]{0.44,0.48,0.49}{##1}}}
\@namedef{PYG@tok@nt}{\let\PYG@bf=\textbf\def\PYG@tc##1{\textcolor[rgb]{0.55,0.00,0.55}{##1}}}
\@namedef{PYG@tok@na}{\def\PYG@tc##1{\textcolor[rgb]{0.40,0.55,0.00}{##1}}}
\@namedef{PYG@tok@nb}{\def\PYG@tc##1{\textcolor[rgb]{0.40,0.55,0.00}{##1}}}
\@namedef{PYG@tok@gh}{\let\PYG@bf=\textbf\def\PYG@tc##1{\textcolor[rgb]{0.00,0.00,0.50}{##1}}}
\@namedef{PYG@tok@gu}{\let\PYG@bf=\textbf\def\PYG@tc##1{\textcolor[rgb]{0.50,0.00,0.50}{##1}}}
\@namedef{PYG@tok@gd}{\def\PYG@tc##1{\textcolor[rgb]{0.67,0.00,0.00}{##1}}}
\@namedef{PYG@tok@gi}{\def\PYG@tc##1{\textcolor[rgb]{0.00,0.67,0.00}{##1}}}
\@namedef{PYG@tok@gr}{\def\PYG@tc##1{\textcolor[rgb]{0.67,0.00,0.00}{##1}}}
\@namedef{PYG@tok@ge}{\let\PYG@it=\textit}
\@namedef{PYG@tok@gs}{\let\PYG@bf=\textbf}
\@namedef{PYG@tok@gp}{\def\PYG@tc##1{\textcolor[rgb]{0.33,0.33,0.33}{##1}}}
\@namedef{PYG@tok@go}{\def\PYG@tc##1{\textcolor[rgb]{0.53,0.53,0.53}{##1}}}
\@namedef{PYG@tok@gt}{\def\PYG@tc##1{\textcolor[rgb]{0.67,0.00,0.00}{##1}}}
\@namedef{PYG@tok@err}{\def\PYG@tc##1{\textcolor[rgb]{0.65,0.09,0.09}{##1}}\def\PYG@bc##1{{\setlength{\fboxsep}{0pt}\colorbox[rgb]{0.89,0.82,0.82}{\strut ##1}}}}
\@namedef{PYG@tok@kc}{\let\PYG@bf=\textbf\def\PYG@tc##1{\textcolor[rgb]{0.55,0.00,0.55}{##1}}}
\@namedef{PYG@tok@kd}{\let\PYG@bf=\textbf\def\PYG@tc##1{\textcolor[rgb]{0.55,0.00,0.55}{##1}}}
\@namedef{PYG@tok@kn}{\let\PYG@bf=\textbf\def\PYG@tc##1{\textcolor[rgb]{0.55,0.00,0.55}{##1}}}
\@namedef{PYG@tok@kp}{\let\PYG@bf=\textbf\def\PYG@tc##1{\textcolor[rgb]{0.55,0.00,0.55}{##1}}}
\@namedef{PYG@tok@kr}{\let\PYG@bf=\textbf\def\PYG@tc##1{\textcolor[rgb]{0.55,0.00,0.55}{##1}}}
\@namedef{PYG@tok@bp}{\def\PYG@tc##1{\textcolor[rgb]{0.40,0.55,0.00}{##1}}}
\@namedef{PYG@tok@fm}{\def\PYG@tc##1{\textcolor[rgb]{0.00,0.55,0.27}{##1}}}
\@namedef{PYG@tok@vc}{\def\PYG@tc##1{\textcolor[rgb]{0.00,0.41,0.55}{##1}}}
\@namedef{PYG@tok@vg}{\def\PYG@tc##1{\textcolor[rgb]{0.00,0.41,0.55}{##1}}}
\@namedef{PYG@tok@vi}{\def\PYG@tc##1{\textcolor[rgb]{0.00,0.41,0.55}{##1}}}
\@namedef{PYG@tok@vm}{\def\PYG@tc##1{\textcolor[rgb]{0.00,0.41,0.55}{##1}}}
\@namedef{PYG@tok@sa}{\def\PYG@tc##1{\textcolor[rgb]{0.80,0.33,0.33}{##1}}}
\@namedef{PYG@tok@sb}{\def\PYG@tc##1{\textcolor[rgb]{0.80,0.33,0.33}{##1}}}
\@namedef{PYG@tok@sc}{\def\PYG@tc##1{\textcolor[rgb]{0.80,0.33,0.33}{##1}}}
\@namedef{PYG@tok@dl}{\def\PYG@tc##1{\textcolor[rgb]{0.80,0.33,0.33}{##1}}}
\@namedef{PYG@tok@sd}{\def\PYG@tc##1{\textcolor[rgb]{0.80,0.33,0.33}{##1}}}
\@namedef{PYG@tok@s2}{\def\PYG@tc##1{\textcolor[rgb]{0.80,0.33,0.33}{##1}}}
\@namedef{PYG@tok@se}{\def\PYG@tc##1{\textcolor[rgb]{0.80,0.33,0.33}{##1}}}
\@namedef{PYG@tok@si}{\def\PYG@tc##1{\textcolor[rgb]{0.80,0.33,0.33}{##1}}}
\@namedef{PYG@tok@s1}{\def\PYG@tc##1{\textcolor[rgb]{0.80,0.33,0.33}{##1}}}
\@namedef{PYG@tok@ss}{\def\PYG@tc##1{\textcolor[rgb]{0.80,0.33,0.33}{##1}}}
\@namedef{PYG@tok@mb}{\def\PYG@tc##1{\textcolor[rgb]{0.71,0.32,0.80}{##1}}}
\@namedef{PYG@tok@mf}{\def\PYG@tc##1{\textcolor[rgb]{0.71,0.32,0.80}{##1}}}
\@namedef{PYG@tok@mh}{\def\PYG@tc##1{\textcolor[rgb]{0.71,0.32,0.80}{##1}}}
\@namedef{PYG@tok@mi}{\def\PYG@tc##1{\textcolor[rgb]{0.71,0.32,0.80}{##1}}}
\@namedef{PYG@tok@il}{\def\PYG@tc##1{\textcolor[rgb]{0.71,0.32,0.80}{##1}}}
\@namedef{PYG@tok@mo}{\def\PYG@tc##1{\textcolor[rgb]{0.71,0.32,0.80}{##1}}}
\@namedef{PYG@tok@ch}{\def\PYG@tc##1{\textcolor[rgb]{0.13,0.55,0.13}{##1}}}
\@namedef{PYG@tok@cm}{\def\PYG@tc##1{\textcolor[rgb]{0.13,0.55,0.13}{##1}}}
\@namedef{PYG@tok@cpf}{\def\PYG@tc##1{\textcolor[rgb]{0.13,0.55,0.13}{##1}}}
\@namedef{PYG@tok@c1}{\def\PYG@tc##1{\textcolor[rgb]{0.13,0.55,0.13}{##1}}}

\def\PYGZus{\char`\_}
\def\PYGZob{\char`\{}
\def\PYGZcb{\char`\}}

\def\PYGZam{\char`\&}

\def\PYGZpc{\char`\%}

\def\PYGZhy{\char`\-}

\def\PYGZdq{\char`\"}
\def\PYGZti{\char`\~}

\makeatother

\begin{document}

\date{}


\title{\Large \bf Is Your Wallet Snitching On You? \\ An Analysis on the Privacy Implications of Web3}

\author{
{\rm Christof Ferreira Torres} \\
ETH Zurich
\and
{\rm Fiona Willi} \\
ETH Zurich
\and
{\rm Shweta Shinde} \\
ETH Zurich
} 

\maketitle
 
\begin{abstract}
With the recent hype around the Metaverse and NFTs, Web3 is getting more and more popular. 
The goal of Web3 is to decentralize the web via decentralized applications. Wallets play a crucial role as they act as an interface between these applications and the user. Wallets such as MetaMask are being used by millions of users nowadays. Unfortunately, Web3 is often advertised as more secure and private. However, decentralized applications as well as wallets are based on traditional technologies, which are not designed with privacy of users in mind. 
 
In this paper, we analyze the privacy implications that Web3 technologies such as decentralized applications and wallets have on users. To this end, we build a framework that measures exposure of wallet information. First, we study whether information about installed wallets is being used to track users online. We analyze the top 100K websites and find evidence of 1,325 websites running scripts that probe whether users have wallets installed in their browser. Second, we measure whether decentralized applications and wallets leak the user's unique wallet address to third-parties. 
We intercept the traffic of 616 decentralized applications and 100 wallets and find over 2000 leaks across 211 applications and more than 300 leaks across 13 wallets. 
Our study shows that Web3 poses a threat to users' privacy and requires new designs towards more privacy-aware wallet architectures.
\end{abstract}

\input{sections/introduction}
\input{sections/background}
\input{sections/methodology}
\input{sections/results}
\input{sections/discussion}
\input{sections/related_work}
\input{sections/conclusion}

\section*{Acknowledgments}


We would like to thank our anonymous reviewers and our shepherd for their valuable comments and feedback. This work was supported by the Zurich Information Security \& Privacy Center (ZISC).

\section*{Availability}


The code that was used to conduct this study as well as the data that was collected during this study is publicly available on GitHub at: \url{https://github.com/christoftorres/Web3-Privacy}.

\bibliographystyle{plain}
\bibliography{references}

\input{sections/appendix}

\end{document}

%% file: sections/introduction.tex
\section{Introduction}

Web3 has gained tremendous adoption over the past few years. This is mainly fueled by the rise of decentralized applications (DApps) such as the Metaverse, NFTs, and decentralized finance (DeFi). 
DappRadar.com currently lists over $13,000$ DApps across various blockchain platforms~\cite{dappradar}.
A report from 2022 states that NFTs generated $12$  Billion USD in trades and that DeFi even reached a value of $127$ Billion USD in total value locked on Ethereum~\cite{dappradar_report}.
The promise of Web3 is the ability to run traditional applications in a decentralized way, thus assuring better transparency and privacy.
An important aspect of such decentralized infrastructure are wallets, which 
act as an interface between decentralized applications and the user. Wallets enable users not only to perform common blockchain operations such as managing their credentials (i.e., public and private key pairs) or signing of transactions, but also operations on DApps such as trading tokens or buying NFTs. All of these operations are provided to the user via a convenient and easy to use interface.
There are several wallet operators that act as intermediaries and interact with decentralized apps on the behalf of the user. 
MetaMask is currently one of the most popular wallet operators with over $10$ Million active users~\cite{metamaskChromeWebStore}.

While built with the goal of better transparency and privacy, decentralized applications as well as wallets are still based on traditional web technologies, which are prone to privacy issues. 
Wallets, in particular have access to sensitive user information and are therefore a rich target for attacks. To make matters worse, wallet operators often use centralized providers by default to retrieve information from the blockchain, making them a single point of failure and allowing providers to easily track user activity across DApps. For example, Infura's recent privacy policy update mentions that IP addresses and wallet addresses of users will be collected \cite{consensys2022policy}. Since Infura is the default blockchain provider of MetaMask, this means that Infura is capable of linking wallet addresses with IP addresses of millions of users.  
%
%
%
While users might accept trusting the wallet operators, they may not realize to what extent they are exposing their wallet information to third-parties.




Wallets operate by injecting a wallet object into the DOM of every website the user visits. This facilitates the interaction between DApps and wallets. DApps can then simply use JavaScript to access wallet information.
However, the browser does not take any particular measures to safeguard the wallet object. Thus, any malicious website, third-party, or browser extension can read this object or use this object to trick users into approving malicious actions (e.g., send assets to an attacker-controlled address). 
While these attack vectors have been exploited in the traditional web in the past, their prevalence in the context of Web3 is yet unclear.

In this paper, we investigate whether wallet extensions are being used to track users online and whether DApps as well as wallet extensions leak the user's wallet address to third-parties. 
To answer this question systematically, we build a framework that is capable of simulating wallet objects and monitoring access to these objects.
Hence, if a website checks the presence of a wallet object in conjunction to several other JavaScript attributes, we deem it as a tracking attempt to fingerprint the user. 
Moreover, our framework is also capable of automatically interacting with DApps as well as wallets and intercepting any cookies as well as requests made via HTTP and WebSockets. We identify a DApp or wallet to leak the user's wallet address if we find that any of the intercepted cookies or requests include the user's wallet address.


\noindent
\newline
\textbf{Results.}
We report three main findings. 
First, of the $100$K websites that we analyzed, $1,325$ of them track users via wallet objects either directly or via third-party scripts. 
Second, of the $1,572$ DApps that we analyzed, $211$ of them leak the user's wallet address to third-parties such as blockchain providers or tracking and analytics platforms.
Lastly, of the $100$ wallets that we analyzed, $13$ wallets leak the user's wallet address to third-parties. All together wallets include over 137 unique third-parties, thereby giving third-parties access to sensitive user information.
In summary, our investigation shows that the existing wallet infrastructure is not in favor of users' privacy. 
Websites are abusing wallets to fingerprint users online, and DApps as well as wallets leak the user's wallet address to third-parties.

\noindent
\newline
\textbf{Ethics Considerations.}
Throughout our analysis, we took adequate measures to avoid overloading the websites (e.g., limited ourselves to the landing page). We have informed the websites and third-parties about potentially unintentional data collection from their side.

\noindent
\newline
\textbf{Contributions.} We summarize our contributions as follows:
\begin{itemize}
    \item We present the first study that systematically measures the prevalence of websites and third-party scripts that use wallet information to track users online. We found evidence that 1,325 websites out of the top 100K websites probe their users for wallets.

    \item We conduct the first large-scale measurement to assess the leakage of wallet addresses on various DApps and wallets. We find that 211 out of 616 DApps and 13 out of 100 wallets leak the user's wallet address to third-parties. 

    \item We measure the efficacy of 5 popular blocklists and observe that when all combined 44\% of the third-parties would not be blocked.
    
\end{itemize}

%% file: sections/background.tex
\section{Background}

We provide background on Ethereum, decentralized applications, wallets, and privacy concerns that might arise when combining all these technologies together.

\subsection{Ethereum}

Ethereum is a blockchain or distributed ledger where transactions are grouped into batches of blocks and where each block points to its previous block via a cryptographic hash. Blockchains are typically maintained by a distributed peer-to-peer network, which is responsible for broadcasting transactions, appending new blocks, providing access to stored data, and executing smart contracts. 
Smart contracts are programs that are deployed and executed across a blockchain. As of January 2023, Ethereum has a market capitalization of over $180$  billion USD~\cite{coinmarketcap}, making it the most popular blockchain technology that offers Turing-complete smart contract capabilities.
Ethereum peers (i.e., nodes) may expose a JSON-RPC interface~\cite{jsonRPCAPI}, which defines an API that users or applications can use to interact with the blockchain (e.g., sending transactions or querying the state of a smart contract).
Similar to other blockchains, Ethereum has its own native cryptocurrency (i.e., Ether), that enables users to transfer value across accounts and to pay for transactions. However, unlike Bitcoin for example, Ethereum follows an account-based model. The idea is similar to traditional bank accounts, where users own an account number and other users may transfer currency to this account number. 
In Ethereum, users do not own an account number, instead they own an account address, which is a unique $160$-bit long hexadecimal string. 
However, similar to a bank account number, addresses act as a unique identifier that can be used to link transactions back to users and which should therefore be shared only with trusted parties.

\subsection{Decentralized Applications}

Decentralized Applications, also known as DApps, are applications that are accessible via the web, but where either all or some of the parts are hosted on decentralized platforms. 
However, for ease-of-use, availability requirements, and compatibility with existing technologies (e.g., DNS, HTTP client-server model, etc.), in most cases the user interface (UI) of DApps is hosted on a centralized web hosting service such as AWS. Only parts of the business logic are decentralized via the use of smart contracts. 
There are a number of different use cases for DApps, ranging from gambling platforms and online games (e.g., CryptoKitties), to decentralized marketplaces and exchanges (e.g., Uniswap). DappRadar currently lists over $3,000$ DApps for the Ethereum blockchain alone~\cite{dappradar}.
However, to be able to interact with DApps, users are required to use a wallet, which acts as a bridge between the DApp and the user's identity on the blockchain.

\begin{figure}[t]
    \centering
    \includegraphics[width=\columnwidth]{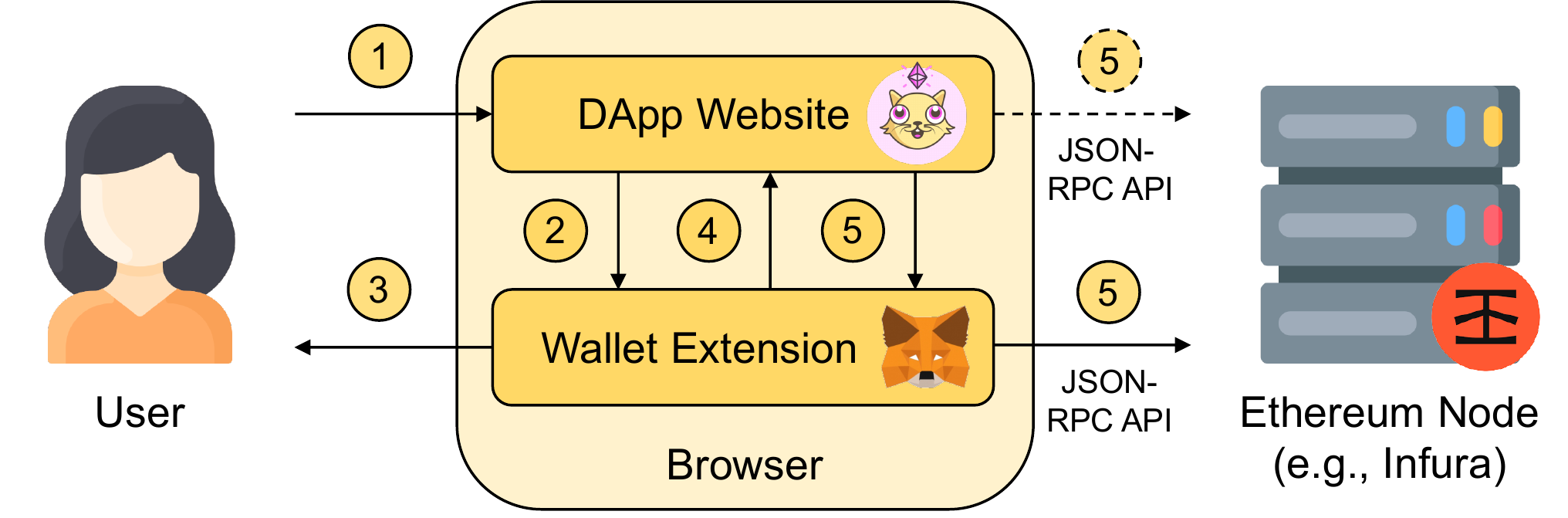}
    \caption{Conceptual flow between a user, a wallet, a DApp and an Ethereum node. \numcircledmod{1} User visits a DApp website and clicks on ``Connect Wallet'' button located on the website, \numcircledmod{2} website requests user's wallet address by calling the wallet extension's Ethereum Provider API~\cite{providerAPI}, \numcircledmod{3} wallet extension asks user for permission and user grants it via the wallet's UI, \numcircledmod{4} wallet returns wallet address to DApp website, \numcircledmod{5} website can now interact with the Ethereum blockchain either directly or through the wallet extension which is connected to an Ethereum node via JSON-RPC API~\cite{jsonRPCAPI}.}
    \label{fig:web3_model}
\end{figure}

\subsection{Wallets}

Users typically manage their accounts and cryptocurrency via a wallet. A popular choice are wallets in the form of a browser extension. A browser extension is a software module that users can install in their browser to enhance their browsing experience. Browser extensions have the capability to modify the Document Object Model (DOM) of websites and enjoy access to privileged browser APIs such as browsing history. 

MetaMask~\cite{metamask} currently is the most popular wallet extension for Ethereum with over $10$ Million downloads on Google Chrome's web store~\cite{metamaskChromeWebStore}. 
Wallet extensions such as MetaMask inject a Web3 object into the DOM of any website that the user visits, regardless of whether the website is a DApp or not. Specifically, MetaMask adds a new object called \texttt{ethereum} to the existing \texttt{window} object, which exposes the Ethereum Provider API~\cite{providerAPI}. The API enables DApps to interact via JavaScript with the Ethereum blockchain as well as the user's wallet. For example, DApps can call unprivileged properties such as \texttt{window.ethereum.isMetaMask}, which will return \texttt{true} if MetaMask is installed, but also privileged properties such as \texttt{window.ethereum.selectedAddress}, which will return the user's wallet address to the DApp. Wallets are required to ask the user for prior permission and the user needs to grant it before a DApp is able to access privileged properties such as the user's wallet address.

\figurename{}~\ref{fig:web3_model} depicts the conceptual flow of a user interacting with a DApp. A user starts by visiting the DApp's website.
DApps usually expose a visual UI button on their website, which users must click if they wish to ``connect'' their wallet to the DApp (i.e., grant DApp access to their wallet). The DApp will then request permission to the wallet via that wallet's injected Ethereum Provider API. 
The wallet will display a popup to the user asking if it wants to grant permission to the DApp. In case the user grants the access, the wallet returns the requested information back to the DApp.
Note that while a subset of the Ethereum Provider API is handled directly by the wallet extension (e.g., signing of transactions), another subset (e.g., retrieving latest block number) is simply forwarded to an Ethereum node (e.g., Infura~\cite{infura}) via JSON-RPC. 
Also note that a DApp is not required to rely on a wallet extension to interact with the blockchain. A DApp can simply talk directly to a blockchain node. In fact, many DApps limit their interaction with the wallet extension to the bare minimum of only requesting the user's wallet address and the signing of transactions. 

\begin{figure*}
    \centering
    \includegraphics[width=0.9\textwidth]{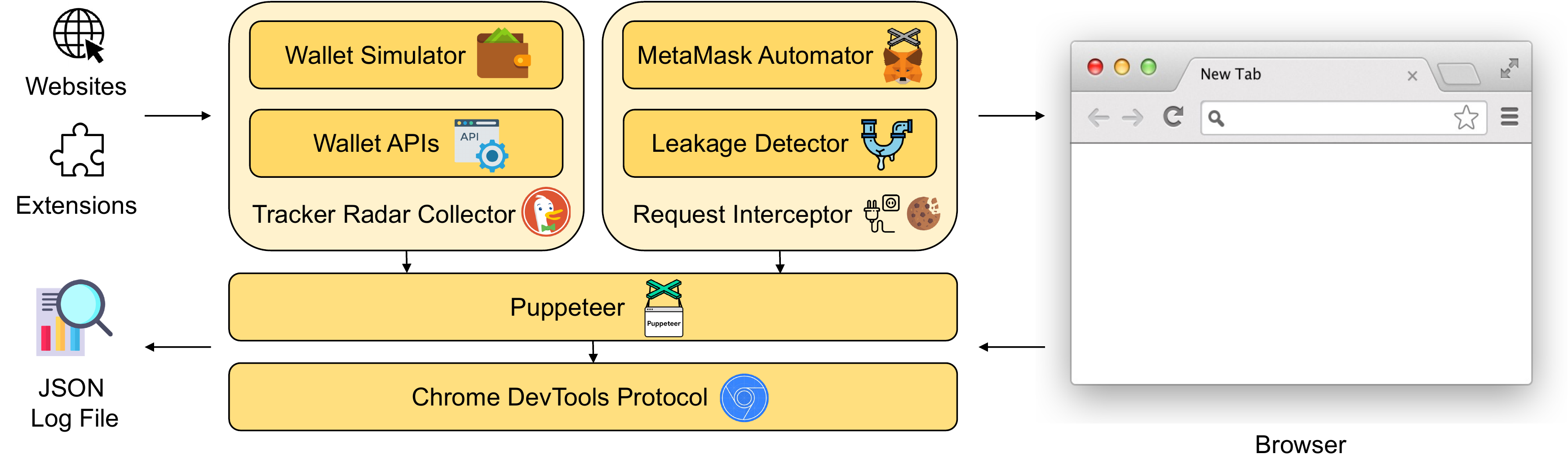}
    \caption{Overview of our measurement framework and its individual components. Our framework takes as input a list of websites or extensions and logs all results in a JSON file.
    We develop a wallet simulator and integrate it into DuckDuckGo's Tracker Radar Collector (TRC)~\cite{trc}. We leverage TRC to crawl websites and log any calls to Wallet APIs. We also develop a Request Interceptor that captures cookies as well as HTTP and WebSocket traffic from websites and extensions. Moreover, we create a MetaMask automator that automates the task of setting up MetaMask~\cite{metamask} as well as connecting to DApps, and integrate a leakage detector to find wallet address leaks. Both, TRC and our Request Interceptor are based on Puppeteer~\cite{puppeteer}, which uses the Chrome DevTools Protocol~\cite{cdp} to interact with a browser.}
    \label{fig:architecture}
\end{figure*}

\subsection{Privacy Concerns}

Tracking is omnipresent on the web. Users are constantly being tracked across websites for purposes of analytics or targeted advertising, either via explicit (e.g., cookies) or implicit (e.g., browser fingerprinting) information. 
In the past, third-party cookies have been a popular way to track users across the web \cite{MayerM12}, but most modern browsers nowadays block third-party cookies by default~\cite{firefox,safari}. A popular alternative is browser fingerprinting \cite{LaperdrixBBA20}. The idea is to uniquely identify users based on differences of their browser's configuration (e.g., fonts, screen resolution, plugins, etc.). A fingerprint is generated by combining properties that are exposed to a website via JavaScript. As opposed to cookies, which are stateful, browser fingerprinting is stateless and thus difficult to mitigate without breaking usability (i.e., disabling JavaScript) \cite{PuglieseRGB20}. 

Since DApps are developed using traditional web technologies, many DApps also include several third-party tracking scripts.
DApps cannot sign transactions without the consent of the user. However, once a user has connected its wallet to a DApp, all third-party scripts embedded within the DApp can access the injected Web3 object via JavaScript. This grants third-party scripts access to sensitive information such as the user's account address or balance, without requiring prior consent of the user. Additionally, without being connected to a wallet, third-party scripts can check for the existence of a Web3 object in the DOM and thus infer that a user owns cryptocurrency and possibly which cryptocurrency and which wallet. This information can be leveraged to augment existing browser fingerprints as it adds additional bits of entropy \cite{Eckersley10}.

While blockchains do provide some level of anonymity (i.e., pseudonymity), they do not provide full anonymity.
All transactions from and to a given account can easily be linked to a user's account address, but not necessarily to its real-world identity. Yet, third-party scripts pose a threat to a user's anonymity since they also have access to a user's IP address and thus can potentially link multiple wallet addresses to their respective IP address \cite{LekiesSWJ15}. In fact, Infura's recent privacy policy update has raised several concerns among the community as it states that wallet addresses and IP addresses will be collected~\cite{consensys2022policy}.

%% file: sections/methodology.tex
\section{Methodology}

Next, we describe our approach for detecting Web3-based browser fingerprinting and identifying wallet address leakage across DApps and wallet extensions. A high-level overview of our measurement framework is depicted in \figurename{}~\ref{fig:architecture}.

\subsection{Web3-Based Browser Fingerprinting}

Browser fingerprinting is a prevalent online tracking technique~\cite{englehardt2016census,VastelLRR18,Gomez-BoixLB18,PuglieseRGB20,LaperdrixBBA20}.
Our goal is to find evidence of whether websites or third-party scripts are leveraging any of the JavaScript properties that wallet extensions expose, to track users on the web.
For that purpose, we use DuckDuckGo's Tracker Radar Collector (TRC)~\cite{trc} to crawl popular websites and measure their behavior. TRC is a crawler that is designed for large-scale web measurements. It is modular and leverages multi-threading to speed up crawling. It uses Puppeteer~\cite{puppeteer} under the hood, which is a library that allows developers to control Chromium-based browsers for automation and testing purposes via the Chrome DevTools Protocol~\cite{cdp}. This gives TRC the capability to intercept network requests, read cookies, and instrument JavaScript calls. 

\subsubsection{Detecting Wallet API Calls}
\label{sec:detecting_wallet_api_calls}

In contrast to OpenWPM~\cite{englehardt2016census}, another popular crawler which uses inline instrumentation by overriding JavaScript functions and objects with getters, TRC uses the Chrome DevTools Protocol to set breakpoints in the JavaScript engine. 
These breakpoints cannot be detected by websites and are only triggered when a certain function is called or property is accessed. 
Whenever the debugger hits any of the configured breakpoints, TRC will collect the JavaScript stack trace (e.g., filename, line number, etc.) and other metadata about the property access or function invocation and store it in a JSON file.

\begin{table}[t]
    \centering
    \begin{adjustbox}{width=\columnwidth,center}
    \begin{tabular}{l l l}
     \toprule
     \textbf{Wallet} & \textbf{TRC Breakpoint} & \textbf{Simulated JavaScript Property} \\
     \midrule 
     MetaMask & \texttt{window.ethereum} & \texttt{isMetaMask: true} \cite{metamaskDocs} \\
     Coinbase & \texttt{window.ethereum} & \texttt{isCoinbaseWallet: true} \cite{coinbaseDocs} \\
     Binance & \texttt{window.BinanceChain} & \texttt{chainId: "0x38"} \cite{binanceDocs} \\
     Phantom & \texttt{window.solana} & \texttt{isPhantom: true} \cite{phantomDocs} \\
     Nami & \texttt{window.cardano} & \texttt{nami.name: "Nami Wallet"} \cite{cardanoDocs} \\
     \bottomrule
    \end{tabular}
    \end{adjustbox}
    \caption{List of breakpoints added to TRC and JavaScript properties simulated by our wallet simulator.}
    \label{tab:wallets}
\end{table}

We set breakpoints for five popular wallets (see \tablename{}~\ref{tab:wallets}). 
We started by first only hooking the \texttt{window.ethereum} object, and added the other wallet API hooks after manually checking reported scripts during initial test runs.
Moreover, after our final crawl we performed a manual inspection of all several scripts and were not able to find any other wallet APIs, which gives us confidence that the four breakpoints in \tableautorefname{} \ref{tab:wallets} are sufficient.
However, breakpoints only get triggered if the object actually exists in the DOM. 
For example, to detect whether websites are trying to identify if MetaMask is installed, we set a breakpoint to be triggered whenever a script accesses the \texttt{window.ethereum} object. 
Thus, in the case of MetaMask, the object \texttt{window.ethereum} has to be injected into the DOM for it to be detected by our breakpoint. We therefore simulate each wallet listed in \tablename{}~\ref{tab:wallets} by injecting (prior to any script execution) wallet specific properties into the DOM. For instance, to simulate MetaMask we inject the property \texttt{window.ethereum.isMetaMask} into the DOM and set it to \texttt{true} as defined in MetaMask's documentation \cite{metamaskDocs}. 
This allows us to hook future accesses by the website, thus catching any access to the object beyond the simulated property. In other words, by hooking \texttt{window.ethereum} via the property \texttt{window.ethereum.isMetaMask}, we will also be able to detect access to, for instance, \texttt{window.ethereum.request}.

\subsubsection{Identifying Fingerprinting Behavior}
\label{sec:identifying_fingerprinting_behavior}

The idea behind browser fingerprinting is to collect a large amount of diverse but stable information about a user's browser configuration, such that when combined together, enough entropy is provided to generate a unique fingerprint that identifies the same user across different sessions and websites. We leverage a similar approach as proposed in \cite{TorresJ18} to detect fingerprinting behavior in Android applications, and adapt it for JavaScript. TRC already provides a curated list of JavaScript properties and functions, that websites are known to leverage, to generate browser fingerprints~\cite{trcAPI}. We group each property and function call into one of $22$ self-defined categories (see Appendix \ref{sec:appendix_a}). A script is marked as a fingerprinting script if it calls JavaScript properties and functions belonging to at least $10$ different categories, where at least one of the categories must belong to a list of $8$ explicit browser fingerprinting categories. We tried values of 5, 10, 15, and 20 during earlier experiments with a small set of identified fingerprinting scripts and achieved the best accuracy when using 10 as threshold.
Explicit browser fingerprinting categories include JavaScript properties and functions that are heavily used for fingerprinting purposes (e.g., CanvasRenderingContext2D, WebGLRenderingContext, AudioBuffer).

\subsection{Wallet Address Leakage}

Nothing prohibits DApps or wallet extensions from sharing a user's wallet address with third-parties. This sharing can either happen with or without the knowledge of the DApp or wallet extension. Our goal is to measure whether, how, and with whom DApps and wallet extensions share wallet addresses. To that end, we developed an automator for MetaMask (see \figurename{}~\ref{fig:architecture}) that not only automatically installs and sets up MetaMask when visiting a DApp, but also automatically tries to connect MetaMask to the Dapp. Once connected to a DApp or a wallet extension is installed, our request interceptor will intercept any outgoing traffic as well as cookies and search for wallet address leaks.

\subsubsection{Connecting MetaMask to DApps}

For a DApp to be able to leak a user's wallet address, the user needs to have set up a wallet and connected its wallet to the DApp. As we do not want to repeat this step manually for thousands of DApps, we develop a component called MetaMask automator. First the  automator sets up MetaMask. This is done even before visiting the website of the DApp. Our automator starts by launching a fresh instance of a browser and installing MetaMask's wallet extension. Afterwards, it launches MetaMask's UI in a new browser page and automatically clicks on the button to import an existing wallet. Browser extensions, including MetaMask, contain UIs that are essentially HTML pages with JavaScript code. Our automator leverages Puppeteer to extract all HTML elements (e.g., buttons, input fields, etc.) from MetaMask's UI. Puppeteer also provides functions that allows our automator to interact with the HTML elements such as clicking on buttons or typing in text into input fields. Once the ``import wallet UI'' has loaded, our automator will read a fixed passphrase and password from a file and automatically type in the passphrase and password into MetaMask's UI to import the wallet and finish setting up MetaMask with a wallet address that we control. 

After setting up the wallet, our automator visits the DApp's website. Once loaded, it searches for a ``connect'' button by scanning the HTML of the DApp's website for elements that contain keywords such as \emph{``Connect Wallet''}, \emph{``Sign In''}, \emph{``Account''}, etc. 
We leveraged the 78 DApps by Winter et al. \cite{winter2021web3} to extract a list of common keywords for connect buttons (see Appendix~\ref{sec:appendix_b}).
Afterwards, the automator tries to perform a click on every element that it found. It detects the connect button if it finds an element where the click succeeds.
Once the automator finds the connect button, it searches for a MetaMask button. Often DApps allow users to connect via different wallets and therefore they let users select which wallet they want to use. The automator finds the MetaMask button by scanning the HTML for elements containing keywords such as \emph{``MetaMask''} or \emph{``Browser Wallet''} (see Appendix~\ref{sec:appendix_b}). Some DApps require users to click on a checkbox to agree to the terms and conditions before being able to connect. Our automator handles this case by searching the HTML for checkboxes and selecting them before clicking on any button. After successfully clicking the MetaMask button, a popup window will show up asking the user for permission to connect. Our automator intercepts this popup window and automatically clicks on the ``confirm'' button to finalize the connection request and give permission to the DApp to access our wallet's information. However, in some cases our automator might not be able to find the MetaMask button via text search because either the DApp uses an image or the text is not detectable. Thus, whenever our automator does not find any MetaMask button, it infers the dimensions of the browser's window and tries to perform hard-coded blind clicks on various offsets starting off from the middle of the window (e.g., $100$ pixels to the bottom right, $50$ pixels to the top left, etc.).

\subsubsection{Intercepting Outgoing Traffic and Cookies}

There are multiple ways in which DApps or wallet extensions can exfiltrate wallet addresses. Previous works have only focused on intercepting HTTP GET requests \cite{winter2021web3}. However, in our work we also intercept HTTP POST requests since DApps and wallet extensions may also leak information via the post body. Moreover, we also intercept WebSocket payloads. WebSockets became a popular alternative to HTTP polling due to their high efficiency (e.g., low latency and fast transmission). They establish a long-lived connection between the DApp or wallet extension and the server. While WebSockets allow for messages to be sent in a bi-directional manner, we are only interested in intercepting outgoing messages (i.e., requests going from the DApp or wallet extension to the server). To that end, we leverage the capabilities of the Chrome DevTools Protocol to intercept network requests to capture any HTTP GET and POST requests as well as outgoing WebSocket messages.
Finally, cookies can also be used to exfiltrate wallet addresses. These can either be set by the server or by the client via JavaScript. We therefore capture cookies that are set via the response headers of HTTP requests and also leverage the capability of the Chrome DevTools Protocol to dump any cookies that were set via JavaScript.

\subsubsection{Identifying Wallet Address Leaks}

We identify wallet address leaks in websites and browser extensions by checking if any of the intercepted traffic (i.e., cookies, HTTP, and WebSockets) contains the wallet address in plain text. More specifically, for cookies, we check whether the value or name of the cookie contains the wallet address. For HTTP GET requests, we check whether the URL of the request contains the wallet address within the GET parameters. For HTTP POST requests, we check whether the post body contains the wallet address. Finally, for WebSockets, we check whether the payload contains the wallet address.
However, checking for the wallet address in plain text is not sufficient. Prior studies~\cite{StarovGN16,EnglehardtHN18,SenolAHB22} have shown that many third-parties often obfuscate their leaks by encoding or hashing them. Identifying obfuscated leaks is a challenging task, which often boils down to a brute-force search. We employ Senol et al.'s~\cite{SenolAHB22} technique, borrowed from Englehardt et al.'s~\cite{EnglehardtHN18} method, to identify email addresses in obfuscated strings. The method consists of searching for a variety of encodings and hashes within strings, by precomputing a set of strings, which contains all possible encodings (e.g., Base64, URL encoding, LZstring, etc.) and hashes (e.g., MD5, SHA256, MurmurHash3, etc.) of the wallet address. Afterwards, the contents of cookies, HTTP requests, and WebSocket payloads are split into multiple strings by potential separator characters (e.g., `=', `\&', etc.) and compared with the strings contained in the precomputed set. This process is repeated until a level of three layers of encodings and decodings is reached.

%% file: sections/results.tex
\section{Measurements}

We describe our experimental setup and present the results of our large-scale measurement to detect web3-based user tracking and wallet address leakage\footnote{Our framework and data are publicly available at: \url{https://github.com/christoftorres/Web3-Privacy}.}.

\subsection{Experimental Setup}

We ran all our experiments on a desktop machine with $10$ cores and $32$GB of RAM. Moreover, we used Chromium version $108.0.5351.0$ as our browser and our automator was build based on MetaMask version $10.22.2$ for Google Chrome.

\noindent
\newline
\textbf{Browser Fingerprinting.}
We measured browser fingerprinting using the top $1$ Million Tranco~\cite{LePochat2019} websites as of November 8th, 2022.\footnote{Available at: \url{https://tranco-list.eu/list/6JXYX/1000000}.} However, Tranco only provides domains and not URLs. Therefore, we tried matching Tranco domains to URLs using Google's Chrome User Experience (CrUX) Report~\cite{crux} of November 2022. Whenever a domain did not match any URL contained in CrUX report, we tried inserting the prefixes \texttt{http(s)://} and \texttt{http(s)://www.} in front of the domains (prioritizing the prefix \texttt{https://www.}) and checked whether these were accessible (i.e., got an HTTP response). We skipped domains that were neither accessible nor contained in the CrUX report.
We started with the top websites (i.e., highest rank to lowest rank) and repeated this process until we had a list of the top $100$K accessible websites. For each website, we limited the maximum crawl duration to $60$ seconds and only visited the landing page.

\begin{table}
    \centering
    \footnotesize
    \begin{tabular}{l r r}
        \toprule
        \textbf{Category} & \textbf{DApps} & \textbf{Valid URLs} \\
        \midrule
        Collectibles & 615 &  533\\
        DeFi & 360 & 339 \\
        Games & 291 & 186 \\
        Other & 220 & 158 \\
        Marketplaces & 97 & 87 \\
        High Risk & 149 & 85 \\
        Exchanges & 87 & 80 \\
        Gambling & 145 & 74 \\
        Social & 34 & 30 \\
        \midrule
        \textbf{Total} & \textbf{1,998} & \textbf{1,572} \\
        \bottomrule
    \end{tabular}
    \caption{DApps crawled from \url{DappRadar.com}.}
    \label{tab:dapp_dataset}
\end{table}

\noindent
\newline
\textbf{Wallet Address Leakage.}
We measured wallet address leakage using three different datasets. The first dataset consists of $66$ DeFi websites from Winter et al.'s study~\cite{winter2021web3}. The second dataset consists of 1,998 DApps that we crawled from DappRadar.com's top Ethereum Dapps~\cite{dappradar}. \tableautorefname{} \ref{tab:dapp_dataset} provides an overview of the number of DApps per category. Note that not all URLs listed on DAppRadar.com are valid. For instance, many URLs in the category collectibles are simply pointing to a collection on \url{opensea.io}. Moreover, some of the URLs are not accessible. We filtered out these URLs and were left with $1,572$ DApps with valid URLs (see \tableautorefname{} \ref{tab:dapp_dataset}), which we then crawled during our experiment. Finally, the third dataset consists of $100$ popular wallet extensions that we downloaded from Google's Chrome Web Store~\cite{webstore}. We installed and set up each wallet extension manually using separate browser profiles for reproducibility. We stored the password and address of each wallet extension in a separate file such that we can afterwards search for the intercepted requests for wallet address leakage.
For each DApp website that we crawled, we limited the maximum crawl duration to $30$ seconds and only visited the landing page. To measure wallet address leakage on wallet extensions, we wrote a script to randomly click on $10$ clickable HTML elements. The interaction with the wallet extension either stops after $10$ elements have been clicked or if $60$ seconds have passed.

\begin{table}[t]
    \centering
    \begin{adjustbox}{width=\columnwidth,center}
    \begin{tabular}{r l l l}
        \toprule
        \textbf{Rank} & \textbf{Website} & \textbf{Script Domain} & \textbf{Wallet API} \\
        \midrule
        74  &  \textbf{tiktok.com}  &  ttwstatic.com  &  All \\
        96  &  \textbf{nytimes.com}  &  googletagmanager.com  &  \texttt{window.ethereum} \\
        160  &  \textbf{xhamster.com}  &  xhcdn.com  &  \texttt{window.ethereum} \\
        & & &  \texttt{window.BinanceChain} \\
        & & &  \texttt{window.solana} \\
        185  &  \textbf{tiktokv.com}  &  ttwstatic.com  &  All  \\
        224  &  \textbf{tinyurl.com}  &  tinyurl.com  &  All \\
        270  &  \textbf{cnbc.com}  &  googlesyndication.com  &  \texttt{window.BinanceChain} \\
        359  &  \textbf{mega.co.nz}  &  mega.io  &  \texttt{window.ethereum} \\
        381  &  \textbf{weather.com}  &  taboola.com  &  \texttt{window.ethereum} \\
        400  &  \textbf{mega.nz}  &  mega.io  &  \texttt{window.ethereum} \\
        481  &  \textbf{pexels.com}  &  pexels.com  &  All \\
        \bottomrule
    \end{tabular}
    \end{adjustbox}
    \caption{Top 10 most ranked websites calling wallet APIs.}
    \label{tab:top_websites}
\end{table}

\begin{table}[t]
    \centering
    \begin{adjustbox}{width=\columnwidth,center}
    \begin{tabular}{l r}
        \toprule
        \textbf{Wallet APIs Combinations} & \textbf{Scripts} \\ 
        \midrule
        \texttt{window.ethereum} 	& 210 \\
        \texttt{window.solana}, \texttt{window.ethereum} 	& 13 \\
        \texttt{window.solana} &	 11 \\
        \texttt{window.ethereum}, \texttt{window.BinanceChain} 	& 8 \\
        \texttt{window.BinanceChain} 	& 5 \\
        \texttt{window.cardano} 	& 4 \\
        \texttt{window.ethereum}, \texttt{window.solana}, \texttt{window.BinanceChain} 	& 3 \\
        \texttt{window.ethereum}, \texttt{window.solana} 	& 2 \\
        \texttt{window.ethereum}, \texttt{window.BinanceChain}, \texttt{window.solana} 	& 2 \\
        \texttt{window.ethereum}, \texttt{window.cardano}, \texttt{window.solana} &	 1 \\
        \texttt{window.BinanceChain}, \texttt{window.ethereum} 	& 1 \\
        \bottomrule
    \end{tabular}
    \end{adjustbox}
    \caption{Observed combinations of explicit wallet API calls.}
    \label{tab:explicit_calls}
\end{table}

\begin{figure} 
    \centering
\begin{footnotesize}
\begin{Verbatim}[frame=single,commandchars=\\\{\}]
\PYG{p}{...}
\PYG{n+nx}{f} \PYG{o}{=} \PYG{n+nb}{Math}\PYG{p}{.}\PYG{n+nx}{trunc}\PYG{p}{((}\PYG{o+ow}{new} \PYG{n+nb}{Date}\PYG{p}{).}\PYG{n+nx}{getTimezoneOffset}\PYG{p}{()} \PYG{o}{/} \PYG{o}{\PYGZhy{}}\PYG{l+m+mf}{60}\PYG{p}{),}
\PYG{n+nx}{m} \PYG{o}{=} \PYG{n+nb}{Boolean}\PYG{p}{(}\PYG{n+nb}{window}\PYG{p}{.}\PYG{n+nx}{web3} \PYG{o}{||} \PYG{n+nb}{window}\PYG{p}{.}\PYG{n+nx}{ethereum}\PYG{p}{),}
\PYG{p}{...}
\PYG{k}{return} \PYG{p}{\PYGZob{}}
  \PYG{p}{...}
  \PYG{n+nx}{isAdBlock}\PYG{o}{:} \PYG{n+nx}{n}\PYG{p}{,}
  \PYG{n+nx}{isMetaMaskActive}\PYG{o}{:} \PYG{n+nx}{m}\PYG{p}{,}
  \PYG{p}{...}
\PYG{p}{\PYGZcb{}}
\PYG{p}{...}
\end{Verbatim}
\end{footnotesize}
    \caption{Code snippet accessing \texttt{window.ethereum} from \url{https://js.wpadmngr.com/static/adManager.m.js}.}
    \label{fig:snippet1}
\end{figure}

\subsection{Web3-Based Browser Fingerprinting}
\label{sec:tracking_results}

\noindent
\newline
\textbf{Wallet API Calls.}
TRC was able to crawl $96,905$ out of $100$K websites successfully (i.e., $96.91\%$).
We found $1,114$ unique scripts on $1,325$ websites which made in total $1,517$ JavaScript calls to at least one wallet APIs listed in \tablename{}~\ref{tab:wallets}. 
\tablename{}~\ref{tab:top_websites} lists the top $10$ most ranked websites which we found to call at least one wallet API. This list includes websites with millions of daily users such as TikTok and the New York Times. 
Interestingly, websites such as TikTok called all of our wallet APIs. After inspecting their code we found that these websites detect whether objects were added to the DOM. We checked whether this only occurs via our wallet simulator or if it also happens when visiting TikTok with MetaMask installed. Our check revealed the same results. This is because MetaMask and any other wallet will, similar to our wallet simulator, inject a new Web3 object into the DOM. This will be detectable by those websites and used for either analytical or tracking purposes. We therefore differentiate between explicit calls and implicit calls, where explicit means that a script includes an explicit call to a wallet API in their code and where implicit means that a script implicitly calls a wallet API when searching for new objects that were added into the DOM. In our experiments, we found that browser fingerprinting scripts (see \tableautorefname{} \ref{tab:third_parties}) often enumerate the entirety of the window object using, for example, \texttt{Object.getOwnPropertyNames(window)} to create a unique fingerprint and thereby implicitly calls a wallet API as it is often part of the window object.
We found $249$ scripts performing explicit calls ($22\%$) and $866$ scripts performing implicit calls ($78\%$). 
\tablename{}~\ref{tab:explicit_calls} lists all combinations that we observed of explicit wallet API calls. We observed in total $11$ combinations, where a simple call to \texttt{window.ethereum} was the most popular call with $210$ scripts calling this wallet API. 

\begin{table*}
    \centering
    \scriptsize
    \begin{tabular}{l r r r r}
        \toprule
        \textbf{Category} & \textbf{Websites} & \textbf{Third-Party Calls} & \textbf{Top Website (Rank)} & \textbf{Top Third-Party (Websites)} \\
        \midrule
        Pornography \& Sexuality  &  200  &  138 (69\%)  &  xhamster.com (160)  &  adsco.re (45) \\
        Computers \& Internet  &  108  &  56 (52\%)  &  tinyurl.com (224)  &  cloudflare.com (10) \\
        News \& Media  &  89  &  65 (73\%)  &  nytimes.com (96)  &  googlesyndication.com (19) \\
        Finances  &  79  &  25 (32\%)  &  opensea.io (1096)  &  cloudflare.com (7) \\
        Adult Sites  &  64  &  46 (72\%)  &  hitomi.la (1066)  &  adsco.re (11) \\
        Entertainment  &  48  &  24 (50\%)  &  bustle.com (2386)  &  googlesyndication.com (7) \\
        E-commerce  &  41  &  16 (39\%)  &  beget.com (1124)  &  cloudflare.com (6) \\
        Business  &  40  &  15 (38\%)  &  bytedance.com (2693)  &  cloudflare.com (6) \\
        Shopping  &  39  &  15 (38\%)  &  moneysavingexpert.com (5419)  &  cloudflare.com (5) \\
        Games  &  38  &  11 (29\%)  &  steamdb.info (4589)  &  m2.ai (2) \\
        \bottomrule
    \end{tabular}
    \caption{Top 10 categories sorted by number of websites detected performing calls to wallet APIs.}
    \label{tab:categories}
\end{table*}

\noindent
\newline
\textbf{Browser Fingerprinting Prevalence.} Following our method defined in Section~\ref{sec:identifying_fingerprinting_behavior} to identify browser fingerprinting, we find that $878$ scripts ($79\%$) belonging to $1,099$ websites ($83\%$) engage in browser fingerprinting and leverage wallet information to enhance the fingerprints they generate. The maximum number of fingerprinting categories collected by a single script was $19$ out of $22$. Both mean and average number of fingerprinting categories collected by browser fingerprinting scripts is around $12$. Also, $71$ ($8\%$) of the scripts performing browser fingerprinting, collected wallet information explicitly whereas $808$ ($92\%$) of the scripts collected wallet information implicitly. \figurename{}~\ref{fig:snippet1} and \figurename{}~\ref{fig:snippet2} list each a small snippet from two third-party scripts that were detected by our framework. Both snippets check for the existence of wallet APIs. \figurename{}~\ref{fig:snippet2} tries to check whether Ethereum, Binance Chain, or Solana wallet extensions are installed and sends this information back to the third-party server via an HTTP POST request.

\begin{figure}
    \centering
\begin{footnotesize}
\begin{Verbatim}[frame=single,commandchars=\\\{\}]
\PYG{n+nb}{document}\PYG{p}{.}\PYG{n+nx}{addEventListener}\PYG{p}{(}\PYG{l+s+s2}{\PYGZdq{}DOMContentLoaded\PYGZdq{}}\PYG{p}{,}
 \PYG{p}{(}\PYG{k+kd}{function}\PYG{p}{()} \PYG{p}{\PYGZob{}}
  \PYG{k+kd}{var} \PYG{n+nx}{e} \PYG{o}{=} \PYG{p}{(}\PYG{l+m+mf}{0}\PYG{p}{,} \PYG{n+nx}{t}\PYG{p}{.}\PYG{n+nx}{getSettings}\PYG{p}{)(),}
      \PYG{n+nx}{n} \PYG{o}{=} \PYG{o+ow}{void} \PYG{l+m+mf}{0} \PYG{o}{!==} \PYG{n+nb}{window}\PYG{p}{.}\PYG{n+nx}{ethereum}\PYG{p}{,}
      \PYG{n+nx}{o} \PYG{o}{=} \PYG{o+ow}{void} \PYG{l+m+mf}{0} \PYG{o}{!==} \PYG{n+nb}{window}\PYG{p}{.}\PYG{n+nx}{BinanceChain}\PYG{p}{,}
      \PYG{n+nx}{a} \PYG{o}{=} \PYG{o+ow}{void} \PYG{l+m+mf}{0} \PYG{o}{!==} \PYG{n+nb}{window}\PYG{p}{.}\PYG{n+nx}{solana}\PYG{p}{;}
  \PYG{p}{...}
  \PYG{k+kd}{var} \PYG{n+nx}{u} \PYG{o}{=} \PYG{o+ow}{new} \PYG{n+nx}{XMLHttpRequest}\PYG{p}{;}
  \PYG{n+nx}{u}\PYG{p}{.}\PYG{n+nx}{open}\PYG{p}{(}\PYG{l+s+s2}{\PYGZdq{}post\PYGZdq{}}\PYG{p}{,} \PYG{l+s+s2}{\PYGZdq{}/x\PYGZhy{}api\PYGZdq{}}\PYG{p}{,} \PYG{o}{!}\PYG{l+m+mf}{0}\PYG{p}{),} \PYG{p}{...,}
  \PYG{n+nx}{u}\PYG{p}{.}\PYG{n+nx}{send}\PYG{p}{(}\PYG{n+nb}{JSON}\PYG{p}{.}\PYG{n+nx}{stringify}\PYG{p}{([\PYGZob{}}
    \PYG{p}{...}
    \PYG{n+nx}{requestData}\PYG{o}{:} \PYG{p}{\PYGZob{}}
      \PYG{n+nx}{model}\PYG{o}{:} \PYG{p}{\PYGZob{}}
        \PYG{p}{...}
        \PYG{n+nx}{key}\PYG{o}{:} \PYG{l+s+s2}{\PYGZdq{}ext\PYGZus{}detection\PYGZdq{}}\PYG{p}{,}
        \PYG{n+nx}{data}\PYG{o}{:} \PYG{p}{\PYGZob{}}
          \PYG{n+nx}{ethereum}\PYG{o}{:} \PYG{n+nx}{n}\PYG{p}{,}
          \PYG{n+nx}{BinanceChain}\PYG{o}{:} \PYG{n+nx}{o}\PYG{p}{,}
          \PYG{n+nx}{solana}\PYG{o}{:} \PYG{n+nx}{a}
        \PYG{p}{\PYGZcb{}}
      \PYG{p}{\PYGZcb{}}
    \PYG{p}{\PYGZcb{}}
  \PYG{p}{\PYGZcb{}]))}
\PYG{p}{\PYGZcb{}))}
\end{Verbatim}
\end{footnotesize}
    \caption{Code snippet hosted at \url{https://static-lvlt.xhcdn.com/xh-shared/js/v1d487c898d.ext-detect.js} accessing \texttt{window.ethereum}, \texttt{window.BinanceChain}, and \texttt{window.solana} wallet APIs to detect whether the user has any of the three wallet extensions installed.}
    \label{fig:snippet2}
\end{figure}

\begin{table*}
    \begin{adjustbox}{width=\textwidth,center}
    \begin{tabular}{l l l c r r}
        \toprule
        \textbf{Third-Party Name} & \textbf{Third-Party Domain} & \textbf{Third-Party Script} & \textbf{Type} & \textbf{Websites} & \textbf{Min. Rank} \\
        \midrule
        -  &  \textbf{wpadmngr.com}  &  https://js.wpadmngr.com/static/adManager.m.js \textbf{(F)}  &  Explicit  &  55  &  1902 \\
        -  &  \textbf{ba0182aa75.com}  &  https://932d007132.ba0182aa75.com/1511a82de1dab2ee0c95006298aa98af.js \textbf{(F)}   &  Explicit  &  39  &  18392 \\
        xHamster  &  \textbf{xhcdn.com}  &  https://static-lvlt.xhcdn.com/xh-shared/js/v1d487c898d.ext-detect.js   &  Explicit  &  23  &  160 \\
        Taboola  &  \textbf{taboola.com}  &  https://cdn.taboola.com/scripts/cwc.es5.js   &  Explicit  &  22  &  381 \\
        Bustle  &  \textbf{bustle.com}  &  https://cdn2.bustle.com/2023/bustle/main-148fdc658d.js   &  Explicit  &  12  &  2386 \\
        Amazon  &  \textbf{cloudfront.net}  &  https://d2vjcex1bx9gzc.cloudfront.net/media/tags/goldfinchfinance.js   &  Explicit  &  6  &  18136 \\
        Google  &  \textbf{googletagmanager.com}  &  https://www.googletagmanager.com/gtm.js?id=GTM-P528B3   &  Explicit  &  5  &  96 \\
        Adshares  &  \textbf{web3ads.net}  &  https://app.web3ads.net/-/view.js   &  Explicit  &  3  &  17540 \\
        Prospect One  &  \textbf{jsdelivr.net}  &  https://cdn.jsdelivr.net/npm/@ledgerhq/connect-kit@1   &  Explicit  &  3  &  14898 \\
        SpookySwap  &  \textbf{spooky.fi}  &  https://spooky.fi/static/js/7.2ec90594.chunk.js   &  Explicit  &  2  &  105612 \\
        \midrule
        Adscore  &  \textbf{adsco.re}  &  https://c.adsco.re/ \textbf{(F)}  &  Implicit  &  111  &  1066 \\
        Cloudflare  &  \textbf{cloudflare.com}  &  https://challenges.cloudflare.com/cdn-cgi/challenge-platform/h/b/orchestrate/chl\_api/v1 \textbf{(F)}  &  Implicit  &  84  &  1114 \\
        Google  &  \textbf{googlesyndication.com}  &  https://pagead2.googlesyndication.com/bg/KJeI0sMyo1Q6mjhDM9mKcjS2IqRt95c1wIDqLysfd0M.js \textbf{(F)}  &  Implicit  &  76  &  270 \\
        CHEQ  &  \textbf{defybrick.com}  &  https://rock.defybrick.com/placement\_invocation?id=65349 \textbf{(F)}  &  Implicit  &  38  &  2923 \\
        MonetizeMore  &  \textbf{m2.ai}  &  https://m2d.m2.ai/v/pg-221207-f8d-nc-434e7f97016ae258bb936353072d000e.js \textbf{(F)}  &  Implicit  &  15  &  6737 \\
        CHEQ  &  \textbf{cheqzone.com}  &  https://ob.cheqzone.com/clicktrue\_invocation.js?id=8911 \textbf{(F)}  &  Implicit  &  13  &  2577 \\
        -  &  \textbf{zfilm-hd.biz}  &  https://go.zfilm-hd.biz/cdn-cgi/challenge-platform/h/b/orchestrate/managed/v1 \textbf{(F)}  &  Implicit  &  9  &  34846 \\
        Anura &  \textbf{anura.io}  &  https://script.anura.io/request.js?instance=3688597576 \textbf{(F)}  &  Implicit  &  8  &  6035 \\
        -  &  \textbf{rageagainstthesoap.com}  &  https://d.rageagainstthesoap.com/clicktrue\_invocation.js?id=11825 \textbf{(F)}  &  Implicit  &  7  &  4042 \\
        ByteDance  &  \textbf{ttwstatic.com}  &  https://sf16-website-login.neutral.ttwstatic.com/obj/tiktok\_web\_login\_static/webmssdk/1.0.0.1/webmssdk.js \textbf{(F)}  &  Implicit  &  5  &  74 \\
        \bottomrule
    \end{tabular}
    \end{adjustbox}
    \footnotesize
    \caption{Top 10 third-parties with explicit (upper half) and implicit (lower half) wallet API calls. \textbf{(F)} Indicates that the third-party script has been flagged by our methodology as a browser fingerprinting script.}
    \label{tab:third_parties}
\end{table*}

\noindent
\newline
\textbf{Categories.} We compared wallet API calls across website categories. \tablename{}~\ref{tab:categories} lists the top $10$ categories in terms of number of websites that access wallet APIs. We used SafeDNS's website categorization service~\cite{safedns} to assign a category to each website. As shown in \tablename{}~\ref{tab:categories}, Pornography \& Sexuality is where we detected the most number of websites (i.e., $200$) accessing wallet information, where the most popular website was \url{xhamster.com} (ranked $160$ in Tranco). Moreover, $69\%$ of the wallet API calls were performed by a third-party script, where \url{adsco.re} is the most widespread third-party with wallet API calls on $45$ different websites. Websites with most third-party calls are in the category News \& Media ($73\%$), whereas websites with least third-party calls are in the category Games ($29\%$).

\noindent
\newline
\textbf{Third-Parties.}
We found $680$ websites (i.e., $51\%$) that include a third-party which calls a wallet API. 
The wallet API calls originate from $324$ third-party scripts which belong to $118$ unique third-party domains.
\tablename{}~\ref{tab:third_parties} lists the top $10$ third-parties for scripts that perform explicit (upper half) and implicit (lower half) wallet API calls. For explicit calls, we find that the third-party domain \url{wpadmngr.com} is the most widespread (embedded in $55$ websites). For implicit calls, we find that the third-party domain \url{adsco.re} is the most widespread (embedded in $111$ websites).

\noindent
\newline
\textbf{URL and Code Similarity.} 
When analyzing the URLs of the $324$ third-party scripts, we noticed that a large number were similar. Several third-party URLs contain the path \texttt{/cdn-cgi/challenge-platform/h/}. We found that these third-parties most likely deploy Cloudflare's Anti-DDoS protection~\cite{cloudflareDDos}, which consists of some JavaScript code that implicitly accesses wallet API information.
We found $127$ (i.e., $39\%$) such Cloudflare third-party scripts.
We also clustered the remaining $197$ third-party scripts based on their code by grouping scripts together which share the exact same JavaScript code. We found $2$ clusters, one including the two third-parties \url{jsdelivr.net} and \url{unpkg.com} and one including the three third-parties \url{6347032d45.com}, \url{wpadmngr.com}, and \url{ba0182aa75.com}. The former third-parties are content delivery networks hosting the $\tt{web3.js}$ library, which is used by several DApps. The other three third-parties are interesting as we do not know who is running them, but we can see from \tablename{}~\ref{tab:third_parties}, that \url{wpadmngr.com} and \url{ba0182aa75.com} are the two most widely deployed third-party scripts calling wallet APIs explicitly. Moreover, as they share the same code, we can infer that they belong to the same organization and that they are together deployed on $94$ different websites.

\begin{figure}
    \centering
    \includegraphics[width=\columnwidth]{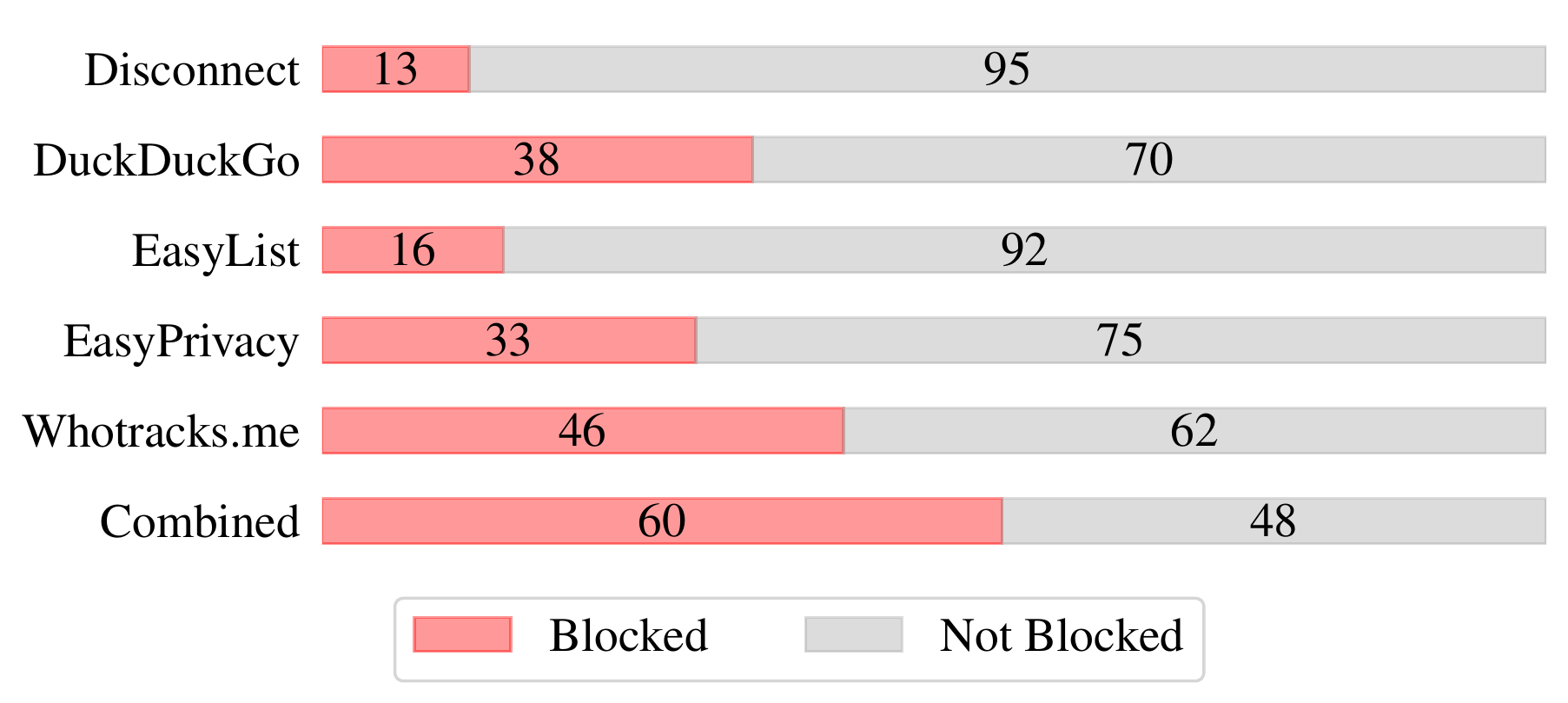}
    \caption{Third-party scripts blocked by popular blocklists.}
    \label{fig:blocklists}
\end{figure}

\noindent
\newline
\textbf{Blocklists.}
Given that half of the calls to wallet APIs originate from third-parties, we checked whether blocklists could be a reliable countermeasure. We downloaded the latest blocklists of Disconnect~\cite{disconnect}, DuckDuckGo~\cite{duckduckgoBlocklist}, EasyList~\cite{easylist}, EasyPrivacy~\cite{easylist}, and Whotracks.me~\cite{whotracksme}, and counted how many of the detected third-parties are blocked by the individual blocklists. 
We manually checked all third-parties and left out 10 of them as they are related to benign use-cases such as helper libraries (e.g., web3.js \cite{web3js}).
\figurename{}~\ref{fig:blocklists} depicts an overview on the number of blocked third-parties.
We observe that Whotracks.me provides the best protection by blocking $46$ third-parties ($43\%$). The weakest protection is given by Disconnect with only 13 third-parties blocked ($12\%$).
Moreover, we also checked whether installing all blocklists at the same time (i.e., combining blocklists) would improve protection. As seen in \figurename{}~\ref{fig:blocklists}, the combination of all five blocklists results in blocking $60$ third-parties ($56\%$), hence an improvement of $12\%$ as compared to only using Whotracks.me's blocklist.

\subsection{Wallet Address Leakage}

We analyze to what extent DApps and wallet extensions leak the user's wallet address to third-parties.

\begin{table}[t]
    \centering
    \begin{adjustbox}{width=\columnwidth,center}
    \begin{tabular}{l r | r r r r}
        \toprule
        \textbf{DeFi Website} & \textbf{GET}~\cite{winter2021web3} & \textbf{GET} & \textbf{POST} & \textbf{WebSockets} & \textbf{Cookies} \\
        \midrule
        \textbf{1inch.io}          & 1  &  0 (0)  &  0 (0)  &  0 (0)  &  0 (0) \\
        \textbf{aave.com}          & 0  &  0 (0)  &  6 (0)  &  0 (0)  &  0 (0) \\
        \textbf{airswap.io}        & 0  &  0 (0)  &  0 (0)  &  1 (0)  &  0 (0) \\
        \textbf{akropolis.io}      & 0  &  0 (0)  &  0 (0)  &  777 (0)  &  0 (0) \\
        \textbf{alchemix.fi}       & 0  &  7 (0)  &  551 (0)  &  0 (0)  &  0 (0) \\
        \textbf{balancer.fi}       & 0  &  0 (0)  &  7 (0)  &  2 (0)  &  0 (0) \\
        \textbf{bancor.network}    & 2  &  1 (0)  &  30 (0)  &  0 (0)  &  0 (0) \\
        \textbf{barnbridge.com}    & 0  &  0 (0)  &  7 (0)  &  0 (0)  &  0 (0) \\
        \textbf{bifi.finance}      & 3  &  3 (3)  &  3 (0)  &  0 (0)  &  0 (0) \\
        \textbf{boringdao.com}     & 0  &  0 (0)  &  6 (0)  &  0 (0)  &  0 (0) \\
        \textbf{centrifuge.io}     & 0  &  0 (0)  &  8 (0)  &  0 (0)  &  0 (0) \\
        \textbf{codefi.network}    & 0  &  0 (0)  &  40 (0)  &  0 (0)  &  0 (0) \\
        \textbf{cream.finance}     & 0  &  1 (0)  &  0 (0)  &  0 (0)  &  0 (0) \\
        \textbf{debank.com}        & 0  &  2 (0)  &  4 (0)  &  0 (0)  &  0 (0) \\
        \textbf{defisaver.com}     & 3  &  2 (2)  &  0 (0)  &  0 (0)  &  0 (0) \\
        \textbf{dmm.exchange}      & 0  &  14 (0)  &  0 (0)  &  0 (0)  &  3 (0) \\
        \textbf{dodoex.io}         & 2  &  0 (0)  &  2 (2)  &  0 (0)  &  0 (0) \\
        \textbf{dydx.exchange}     & 0  &  0 (0)  &  43 (0)  &  0 (0)  &  0 (0) \\
        \textbf{enzyme.finance}    & 0  &  0 (0)  &  5 (0)  &  0 (0)  &  0 (0) \\
        \textbf{fei.money}         & 0  &  0 (0)  &  7 (0)  &  0 (0)  &  0 (0) \\
        \textbf{foundation.app}    & 0  &  0 (0)  &  6 (0)  &  0 (0)  &  0 (0) \\
        \textbf{idle.finance}      & 0  &  1 (0)  &  3 (0)  &  0 (0)  &  0 (0) \\
        \textbf{impermax.finance}  & 1  &  0 (0)  &  0 (0)  &  0 (0)  &  0 (0) \\
        \textbf{indexcoop.com}     & 0  &  0 (0)  &  55 (0)  &  0 (0)  &  0 (0) \\
        \textbf{inverse.finance}   & 0  &  0 (0)  &  2 (0)  &  0 (0)  &  0 (0) \\
        \textbf{jelly.market}      & 1  &  0 (0)  &  0 (0)  &  0 (0)  &  0 (0) \\
        \textbf{liquity.app}       & 0  &  0 (0)  &  0 (0)  &  16 (0)  &  0 (0) \\
        \textbf{mai.finance}       & 0  &  0 (0)  &  53 (0)  &  0 (0)  &  0 (0) \\
        \textbf{notional.finance}  & 0  &  0 (0)  &  27 (0)  &  6 (0)  &  0 (0) \\
        \textbf{o3swap.com}        & 0  &  0 (0)  &  70 (0)  &  0 (0)  &  0 (0) \\
        \textbf{opensea.io}        & 0  &  0 (0)  &  8 (0)  &  0 (0)  &  0 (0) \\
        \textbf{opyn.co}           & 0  &  0 (0)  &  4 (0)  &  2 (0)  &  0 (0) \\
        \textbf{pickle.finance}    & 0  &  0 (0)  &  258 (0)  &  0 (0)  &  0 (0) \\
        \textbf{rari.capital}      & 0  &  0 (0)  &  16 (0)  &  0 (0)  &  0 (0) \\
        \textbf{rarible.com}       & 3  &  0 (0)  &  1 (0)  &  0 (0)  &  0 (0) \\
        \textbf{reflexer.finance}  & 1  &  2 (2)  &  1 (0)  &  0 (0)  &  0 (0) \\
        \textbf{sablier.finance}   & 1  &  0 (0)  &  6 (3)  &  0 (0)  &  0 (0) \\
        \textbf{truefi.io}         & 0  &  0 (0)  &  8 (0)  &  0 (0)  &  0 (0) \\
        \textbf{uniswap.org}       & 0  &  0 (0)  &  4 (0)  &  0 (0)  &  0 (0) \\
        \textbf{warp.finance}      & 0  &  0 (0)  &  2 (0)  &  0 (0)  &  0 (0) \\
        \textbf{yearn.finance}     & 4  &  0 (0)  &  0 (0)  &  0 (0)  &  0 (0) \\
        \textbf{yield.is}          & 0  &  0 (0)  &  65 (0)  &  0 (0)  &  0 (0) \\
        \textbf{zerion.io}         & 3  &  0 (0)  &  16 (0)  &  0 (0)  &  0 (0) \\
        \midrule
        \textbf{Total}             & 25  &  33 (7)  &  1324 (5)  &  804 (0)  &  3 (0) \\
        \bottomrule
    \end{tabular}
    \end{adjustbox}
    \caption{Leaks measured by Winter et al.~\cite{winter2021web3} vs our framework. Numbers in parentheses indicate common leaks.}
    \label{tab:winter_leaks}
\end{table}

\subsubsection{DApps}

\begin{table*}
    \centering
    \scriptsize
    \begin{tabular}{r l l l c r | r r r r}
        \toprule
        \textbf{\#} & \textbf{Third-Party Name} & \textbf{Third-Party Domain} & \textbf{Category} & \textbf{Collects IP Address} & \textbf{DApps} & \textbf{GET} & \textbf{POST} & \textbf{WebSockets} & \textbf{Cookies} \\
        \midrule
        1 & Infura &  \textbf{infura.io}          & JSON-RPC Provider  & Yes  &  42  &  0  &  818  &  12  &  0 \\
        2 & Alchemy &  \textbf{alchemyapi.io}      & JSON-RPC Provider  &  Yes &  39  &  0  &  638  &  772  &  0 \\
        3 & The Graph &  \textbf{thegraph.com}       & JSON-RPC Provider   & Yes  &  22  &  0  &  230  &  0  &  0 \\
        4 & Sentry &  \textbf{sentry.io}          & Tracking \& Analytics   & Yes  &  21  &  0  &  122  &  0  &  0 \\
        5 & Google &  \textbf{google-analytics.com}  & Tracking \& Analytics   & Yes  &  18  &  38  &  6  &  0  &  0 \\
        6 & Alchemy &  \textbf{alchemy.com}        & JSON-RPC Provider  & Yes  &  16  &  3  &  132  &  6  &  0 \\
        7 & Amplitude &  \textbf{amplitude.com}      & Tracking \& Analytics   & Yes  &  13  &  0  &  37  &  0  &  0 \\
        8 & Blocknative &  \textbf{blocknative.com}    & JSON-RPC Provider  & Yes  &  12  &  0  &  0  &  25  &  0 \\
        9 & Ankr &  \textbf{ftm.tools}          & JSON-RPC Provider  & Yes  &  11  &  0  &  166  &  0  &  0 \\
        10 & Binance &  \textbf{binance.org}        &  JSON-RPC Provider &  Yes &  10  &  0  &  99  &  0  &  0 \\
        11 & Ankr &  \textbf{ankr.com}           & JSON-RPC Provider  & Yes  &  9  &  0  &  316  &  0  &  0 \\
        12 & Mixpanel &  \textbf{mixpanel.com}       & Tracking \& Analytics  & Yes  &  9  &  17  &  26  &  0  &  0 \\
        13 & Etherscan &  \textbf{etherscan.io}       &  JSON-RPC Provider & Yes  &  8  &  38  &  0  &  0  &  0 \\
        14 & Binance &  \textbf{ninicoin.io}        & JSON-RPC Provider  & Yes  &  7  &  0  &  52  &  0  &  0 \\
        15 & Arbitrum &  \textbf{arbitrum.io}        & JSON-RPC Provider  &  Yes &  6  &  0  &  67  &  0  &  0 \\
        16 & Avalanche &  \textbf{avax.network}       & JSON-RPC Provider  & Yes  &  6  &  0  &  47  &  0  &  0 \\
        17 & Cloudflare &  \textbf{cloudflare-eth.com}  & JSON-RPC Provider  &  Yes &  6  &  0  &  19  &  0  &  0 \\
        18 & Google &  \textbf{firestore.googleapis.com}  & Tracking \& Analytics  & Yes  &  6  &  22  &  17  &  0  &  0 \\
        19 & Pocket Network &  \textbf{pokt.network}       & JSON-RPC Provider  & No  &  6  &  0  &  25  &  0  &  0 \\
        20 & Ankr &  \textbf{polygon-rpc.com}    & JSON-RPC Provider  & Yes  &  6  &  0  &  22  &  0  &  0 \\
        \bottomrule
    \end{tabular}
    \caption{Top $20$ third-parties detected by our framework on the DappRadar.com to which the wallet address was leaked.}
    \label{tab:dapp_thirdparties}
\end{table*}

\textbf{Winter et al.~\cite{winter2021web3}.} We compare the performance of our framework using Winter et al.'s~\cite{winter2021web3} DeFi dataset. The dataset consists of $78$ DeFi websites, however, $6$ websites were down at the time of writing, $2$ websites did not support MetaMask, and $4$ were duplicates. After filtering, we were left with $66$ websites to crawl. While Winter et al.~connect manually to each website via MetaMask, we automatically connect to all of them using our MetaMask automator.
\tablename{}~\ref{tab:winter_leaks} shows a comparison between the leaks measured by Winter et al.~and our framework.
Winter et al.~found that $13$ out of the $66$ websites ($20\%$), leak the user's wallet address to a third-party, whereas our results show that actually $39$ out of the 66 websites ($59\%$) leak the user's wallet address to a third-party.
Overall, Winter et al.~found 25 leaks whereas we found $2,164$ leaks for the same websites. $98\%$ (i.e., $2,131$ leaks) are performed either via POST requests, WebSockets, or cookies. $61\%$ of the leaks (i.e., $1,324$ leaks) occur via POST requests. This emphasizes that solely analyzing GET requests, as Winter et al.~did, is not sufficient.
While Winter et al.~found that wallet addresses are being leaked to $14$ third-parties, our results show that the actual number is much higher, namely $64$ third-parties.
\tablename{}~\ref{tab:winter_leaks} highlights leaks that our framework and Winter et al.~have in common (number between parentheses). For example, for \url{bifi.finance}, we detected 3 leaks which correspond to the same leaks as detected by Winter et al. 
However, we observe that \url{1inch.io}, \url{impermax.finance}, \url{jelly.market}, and \url{yearn.finance} did not leak the user's wallet address in our crawls anymore. On closer inspection, we find that \url{jelly.market} was down and we therefore were not able to collect any data and that the remaining three websites moved towards using their own API to retrieve blockchain data.
Interestingly, for \url{dodoex.io} and \url{sablier.finance}, the leaks moved from GET requests to POST requests. This can be due to a change in the API of the third-parties. 
Moreover, while \url{alchemix.fi}, \url{cream.finance}, \url{debank.com}, \url{dmm.exchange}, and \url{idle.finance} did not leak the user's wallet address to any third-parties via GET requests during Winter et al.'s study, our results demonstrate the opposite. Since Winter et al.'s study is already more than a year ago, we assume that those third-party leaks were added after the study was conducted.
Finally, we also observed that \url{dmm.exchange} leaks the user's wallet address to \url{kyberswap.com} via 3 different cookies set by Mixpanel (see \figureautorefname{} \ref{fig:cookie} for an example of such a cookie).

\begin{figure}
    \centering
\begin{footnotesize}
\begin{Verbatim}[frame=single,commandchars=\\\{\},codes={\catcode`\$=3\catcode`\^=7\catcode`\_=8\relax}]
\PYG{p}{\PYGZob{}}
  \PYG{l+s+s2}{\PYGZdq{}name\PYGZdq{}}\PYG{o}{:} \PYG{l+s+s2}{\PYGZdq{}mp\PYGZus{}ff1eea26c19dcf4a7c35ebbc8631e714\PYGZus{}mixpanel\PYGZdq{}}\PYG{p}{,}
  \PYG{l+s+s2}{\PYGZdq{}value\PYGZdq{}}\PYG{o}{:} \PYG{l+s+s2}{\PYGZdq{}\PYGZpc{}7B\PYGZpc{}22distinct\PYGZus{}id\PYGZpc{}22\PYGZpc{}3A\PYGZpc{}20\PYGZpc{}220x7e4ABd63A7C8}
\PYG{l+s+s2}{  314Cc28D388303472353D884f292\PYGZpc{}22\PYGZpc{}2C\PYGZpc{}22\PYGZpc{}24device\PYGZus{}id\PYGZpc{}22\PYGZpc{}}
\PYG{l+s+s2}{  3A\PYGZpc{}20\PYGZpc{}22185bc157265a0d\PYGZhy{}0daab5a6ab23c7\PYGZhy{}17525635\PYGZhy{}16a7f0}
\PYG{l+s+s2}{  \PYGZhy{}185bc157266f56\PYGZpc{}22\PYGZpc{}2C\PYGZpc{}22\PYGZpc{}24user\PYGZus{}id\PYGZpc{}22\PYGZpc{}3A\PYGZpc{}20\PYGZpc{}220x7e4AB}
\PYG{l+s+s2}{  d63A7C8314Cc28D388303472353D884f292\PYGZpc{}22\PYGZpc{}2C\PYGZpc{}22\PYGZpc{}24initia}
\PYG{l+s+s2}{  l\PYGZus{}referrer\PYGZpc{}22\PYGZpc{}3A\PYGZpc{}20\PYGZpc{}22\PYGZpc{}24direct\PYGZpc{}22\PYGZpc{}2C\PYGZpc{}22\PYGZpc{}24initial\PYGZus{}re}
\PYG{l+s+s2}{  ferring\PYGZus{}domain\PYGZpc{}22\PYGZpc{}3A\PYGZpc{}20\PYGZpc{}22\PYGZpc{}24direct\PYGZpc{}22\PYGZpc{}2C\PYGZpc{}22wallet\PYGZus{}ad}
\PYG{l+s+s2}{  dress\PYGZpc{}22\PYGZpc{}3A\PYGZpc{}20\PYGZpc{}220x7e4ABd63A7C8314Cc28D388303472353D8}
\PYG{l+s+s2}{  84f292\PYGZpc{}22\PYGZpc{}2C\PYGZpc{}22platform\PYGZpc{}22\PYGZpc{}3A\PYGZpc{}20\PYGZpc{}22Web\PYGZpc{}22\PYGZpc{}2C\PYGZpc{}22networ}
\PYG{l+s+s2}{  k\PYGZpc{}22\PYGZpc{}3A\PYGZpc{}20\PYGZpc{}22Ethereum\PYGZpc{}22\PYGZpc{}7D\PYGZdq{}}\PYG{p}{,}
  \PYG{l+s+s2}{\PYGZdq{}domain\PYGZdq{}}\PYG{o}{:} \PYG{l+s+s2}{\PYGZdq{}.kyberswap.com\PYGZdq{}}\PYG{p}{,}
  \PYG{p}{...}
\PYG{p}{\PYGZcb{}}
\end{Verbatim}
\end{footnotesize}
    \caption{Cookie set on \url{dmm.exchange} by Mixpanel for \url{.kyberswap.com} domain containing user's wallet address: \textcolor{red}{\texttt{\small0x7e4ABd63A7C8314Cc28D388303472353D884f292}}.}
    \label{fig:cookie}
\end{figure}

\noindent
\newline
\textbf{DappRadar.com~\cite{dappradar}.} Winter et al.'s dataset is useful for comparing performance, but it is insufficient to draw any general conclusions due to it being relatively small and only focusing on DeFi websites. Therefore, we crawled DappRadar.com to obtain a much larger and diverse dataset. We ended up getting $1,572$ DApp websites across $9$ categories. Our automator was able to automatically connect to $616$ ($39\%$) of them. The automator had less issues in connecting to DeFi DApps, with a success rate of $61\%$. On the other hand, our automator found it hard to connect to High Risk DApps, with a success rate of only $20\%$. 
There are several reasons why it was not able to connect to all DApps. Most websites are simply down or our automator is not able to detect the connect button by scanning the HTML. Some websites do not support MetaMask, or require users to either agree on the terms or register and login via an email address and a password before being able to interact with the DApp. Section \ref{sec:limitations} provides a clear breakdown regarding why our automator was not able to connect to certain DApps.

\tablename{}~\ref{tab:dappradar_results} summarizes our detected leaks on the \url{DappRadar.com} dataset. Our framework identified $211$ unique DApp websites ($35\%$ of the connected DApps) which leak the user's wallet address across $137$ unique third-parties. 
As shown in \tablename{}~\ref{tab:dappradar_results}, Gambling DApps leak the least ($6\%$), whereas Exchanges leak the most ($59\%$).
Our data also shows that $1,400$ DApps ($89\%$) embed at least one third-party. On average DApps embed $7$ different third-parties. The maximum we observed was $61$ third-parties embedded on a single DApp's website. 
\tablename{}~\ref{tab:dapp_thirdparties} lists the top $20$ third-parties where the user's wallet address is leaked to.
As we can see, most third-parties are JSON-RPC providers ($75\%$) and the rest are tracking \& analytics platforms ($25\%$).
DApps need to connect to a blockchain node to retrieve blockchain related information. This connection is often performed via JSON-RPC providers. While leaks to JSON-RPC providers are unavoidable, they still may pose a threat to user's privacy as they may collect additional information such as what DApps the user visited or its IP address. Often users do not know to which JSON-RPC provider the DApp is connected to. Leaks to tracking \& analytics platforms are unnecessary and a clear privacy violation. These platforms should not have access to sensible user information such as wallet addresses.
For example, \figureautorefname{} \ref{fig:google_leak} shows an HTTP GET request from \url{degens.farm} that leaks the user's wallet address to \url{google-analytics.com}.
We studied the privacy policies of the top $20$ third-parties and observe that $95\%$ state that they collect the user's IP address. Pocket Network is the only third-party in \tablename{}~\ref{tab:dapp_thirdparties} that does not collect the IP address of its users. 
We also observe that Infura is the most widespread third-party, with $42$ DApps leaking the user's wallet address to Infura.
For the DappRadar.com dataset, none of the DApps shared the user's wallet address via cookies. However, similar to Winter et al.'s dataset, most DApps share the user's wallet address via HTTP POST requests, then WebSockets, and finally HTTP GET requests. 

\begin{figure*}
    \centering
\begin{footnotesize}
\begin{Verbatim}[frame=single,commandchars=\\\{\},codes={\catcode`\$=3\catcode`\^=7\catcode`\_=8\relax}]
 https://www.google\PYGZhy{}analytics.com/collect?v=1\PYGZam{}\PYGZus{}v=j99\PYGZam{}a=1044933369\PYGZam{}t=event\PYGZam{}ni=0\PYGZam{}\PYGZus{}s=1\PYGZam{}dl=https\PYGZpc{}3A\PYGZpc{}2F\PYGZpc{}2Fdegens.farm\PYGZpc{}2Fwallet\PYGZam{}
 ul=en\PYGZhy{}us\PYGZam{}de=UTF\PYGZhy{}8\PYGZam{}dt=Degen\PYGZpc{}27\PYGZpc{}24\PYGZpc{}20Farm\PYGZpc{}3A\PYGZpc{}20Wallet\PYGZam{}sd=30\PYGZhy{}bit\PYGZam{}sr=1512x982\PYGZam{}vp=1512x749\PYGZam{}je=0\PYGZam{}ec=WalletConnected\PYGZam{}ea=\PYG{esc}{\textcolor{red}{0x7e4abd}}
 \PYG{esc}{\textcolor{red}{63a7c8314cc28d388303472353d884f292}}\PYGZam{}el=labelForWalletConnect\PYGZam{}ev=7.20999590401511e\PYGZpc{}2B47\PYGZam{}\PYGZus{}u=aADAAEABAAAAACAAI\PYGZti{}\PYGZam{}jid=\PYGZam{}gjid=\PYGZam{}ci
 d=437541385.1675387202\PYGZam{}tid=UA\PYGZhy{}201259489\PYGZhy{}1\PYGZam{}\PYGZus{}gid=196110690.1675387203\PYGZam{}gtm=2wg2105PC69BZ\PYGZam{}z=1330733511
\end{Verbatim}
\end{footnotesize}
    \caption{Wallet address leaked via HTTP GET request to \url{google-analytics.com} on the \url{degens.farm} DApp.}
    \label{fig:google_leak}
\end{figure*}

\begin{table} 
    \centering
    \begin{adjustbox}{width=\columnwidth,center}
    \begin{tabular}{l r r | r r r r}
        \toprule
        \textbf{Category} & \textbf{DApps} & \textbf{Third-Parties} & \textbf{GET} & \textbf{POST} & \textbf{WebSockets} & \textbf{Cookies} \\
        \midrule
        Collectibles  &  32 (20\%)  &  23 (4\%)  &  38  &  77  &  9  &  0 \\
        DeFi  &  93 (45\%)  &  87 (17\%)  &  319  &  2533  &  807  &  0 \\
        Games  &  22 (30\%)  &  20 (6\%)  &  43  &  95  &  6  &  0 \\
        Other  &  21 (35\%)  &  22 (6\%)  &  32  &  227  &  2  &  0 \\
        Marketplaces  &  21 (45\%)  &  25 (8\%)  &  7  &  102  &  4  &  0 \\
        High Risk  &  3 (18\%)  &  5 (4\%)  &  6  &  9  &  0  &  0 \\
        Exchanges  &  19 (59\%)  &  42 (18\%)  &  46  &  574  &  38  &  0 \\
        Gambling  &  1 (6\%)  &  1 (0\%)  &  0  &  1  &  0  &  0 \\
        Social  &  3 (37\%)  &  5 (4\%)  &  4  &  25  &  0  &  0 \\
        \midrule
        \textbf{Total Unique}  &  211 (35\%)  &  137 (9\%)  &  495  &  3643  &  866  &  0 \\
        \bottomrule
    \end{tabular}
    \end{adjustbox}
    \caption{Leaks identified on the \url{DappRadar.com} dataset.}
    \label{tab:dappradar_results}
\end{table}

\begin{table}[h]
    \centering
    \begin{adjustbox}{width=\columnwidth,center}
    \begin{tabular}{l l | r r r r}
    \toprule
    \textbf{Wallet Extension} & \textbf{Third-Party} & \textbf{GET} & \textbf{POST} & \textbf{WebSockets} & \textbf{Cookies} \\
    \midrule
    NuFi  &  \textbf{milkomeda.com (R)}  &  0  &  4  &  0  &  0 \\
    Petra Aptos Wallet  &  \textbf{segment.io}  &  0  &  22  &  0  &  0 \\
    Pitaka  &  \textbf{sentry.io}  &  0  &  2  &  0  &  0 \\
    Nabox Wallet  &  \textbf{mytokenpocket.vip}  &  0  &  11  &  0  &  0 \\
    GameStop Wallet  &  \textbf{loopring.network}  &  97  &  0  &  0  &  0 \\
     &  \textbf{immutable.com}  &  116  &  0  &  0  &  0 \\
    Martian Wallet  &  \textbf{dialectapi.to}  &  16  &  2  &  0  &  0 \\
    Crust Wallet  &  \textbf{subscan.io}  &  0  &  17  &  0  &  0 \\
    JulWallet  &  \textbf{swapliquidity.org}  &  0  &  1  &  0  &  0 \\
    Ethos Sui Wallet  &  \textbf{shinami.com (R)}  &  0  &  16  &  0  &  0 \\
    G.U. Smart Wallet  &  \textbf{infura.io (R)}  &  0  &  1  &  0  &  0 \\
    Coinbase Wallet  &  \textbf{fantom.network}  &  0  &  1  &  0  &  0 \\
     &  \textbf{blockscout.com}  &  2  &  0  &  0  &  0 \\
     &  \textbf{binance.org}  &  0  &  1  &  0  &  0 \\
     &  \textbf{etherscan.io}  &  4  &  0  &  0  &  0 \\
     &  \textbf{avax-test.network}  &  0  &  1  &  0  &  0 \\
     &  \textbf{bscscan.com}  &  4  &  0  &  0  &  0 \\
     &  \textbf{snowtrace.io}  &  4  &  0  &  0  &  0 \\
     &  \textbf{arbiscan.io}  &  4  &  0  &  0  &  0 \\
     &  \textbf{polygonscan.com}  &  4  &  0  &  0  &  0 \\
     &  \textbf{ftmscan.com}  &  4  &  0  &  0  &  0 \\
    Ethereum AllInOne  &  \textbf{ethplorer.io}  &  21  &  0  &  0  &  0 \\
    Verto  &  \textbf{pulsechain.com (R)}  &  0  &  3  &  0  &  0 \\
     &  \textbf{volentix.io}  &  25  &  1  &  0  &  0 \\
    \midrule
    \textbf{Total} &  \textbf{24}  &  301  &  83  &  0  &  0 \\
    \bottomrule
    \end{tabular}
    \end{adjustbox}
    \caption{Third-party leaks detected in $100$ wallet extensions. (R) stands for JSON-RPC provider.}
    \label{tab:extension_leaks}
\end{table}

\subsubsection{Wallet Extensions}

We analyzed whether any of the $100$ wallet extensions contained in our dataset include third-parties and with whom they share the user's wallet address and potentially even the password or browsing history. Fortunately, none of the analyzed browser extensions seem to leak the user's password. At least we were not able to identify the password in any of the requests that we analyzed (including requests to the first-parties themselves). 
We analyzed the manifest file of each wallet extension and checked whether they can inject content scripts on any website and if they request access to sensible permissions. 89 of the 100 wallet extensions can inject content scripts on any website (i.e., the manifest includes one of the following patterns: 'http://*/*', https://*/*',<all\_urls>', *://*/*'). Hence, these 89 wallet could potentially read the URL of the current page and send it to a backend. We also found that 66 wallet extensions request permission for either accessing “history”, “tabs”, or “activeTab”. We visited three different websites (\url{nytimes.com}, \url{etherscan.io}, and \url{uniswap.org}), using each extension and checked whether there are requests that include any of the three websites. We were not able to detect any extension leaking any of the visited websites.
However, we did find that wallet extensions do leak the user's wallet address to third-parties.
\tablename{}~\ref{tab:extension_leaks} lists the wallet extension that leak the user's wallet address. 
We found $13$ out of $100$ analyzed extensions which leak the user's wallet address to at least one of $24$ third-parties.
In total we found $139$ third-parties across all browser extensions.
While most wallet extensions only leak the wallet address to a single third-party, Coinbase's wallet extension leaks the user's wallet address to $10$ different third-parties. Surprisingly, none of the wallet extensions' third-parties seem to overlap. However, we do observe that \url{sentry.io} and \url{infura.io} are present in both of our datasets, DApps and wallet extensions. Although, Infura is a benign platform, since it is a JSON-RPC provider and therefore required for the wallet extension work, the user is not made aware of this connection and the fact that Infura can link requests across websites and infer for example that a user $X$ uses wallet $Y$ and visits DApps $A$ and $B$ regularly. Sentry on the other hand is clearly not benign as it is a tracking \& analytics platform, hence sensitive information such as the user's wallet address should not be leaked to such platforms.

%% file: sections/discussion.tex
\section{Discussion}

We discuss the limitations of our methodology and elaborate potential countermeasures, including their pitfalls.

\subsection{Limitations}
\label{sec:limitations}

Our methodology for detecting wallet API calls is based on TRC which comes built in with an anti-bot detection. However, anti-bot detection solutions are not perfect and thus websites can still detect whether a bot is crawling them and thus behave differently or block access to the website. Moreover, our methodology leverages a wallet simulator that we build to inject fake JavaScript objects into the DOM such that we can simulate wallets without requiring to install them and setting them up. However, our simulator does not simulate a full-fledged wallet. It is limited to the simulated JavaScript properties listed in \tablename{}~\ref{tab:wallets} in Section \ref{sec:detecting_wallet_api_calls}. Thus, third-party scripts could detect our wallet simulator by checking for inconsistencies such as missing JavaScript properties in the different wallet APIs.
Although, the likelihood that third-party scripts currently do this is rather low. We did not experience such checks when analyzing the code of third-party scripts manually. However, it could be that in the future third-party scripts will adapt and try to probe for multiple properties of a wallet before making any decisions. 

Our MetaMask automator was only able to automatically connect to $39\%$ of the analyzed DApps. 
Appendix \ref{sec:appendix_c}, provides a detailed breakdown on connection failures that occurred over the DeFi subset of the DappRadar.com dataset. In 24\% of the cases, the URLs did not point to a valid DApp and in 14\% of the cases the DApp's website was simply down.
3\% of the DApps did not support MetaMask. For 18\% of the DApps our automator could simply not detect a connect button or MetaMask button within the HTML despite the DApp's buttons containing labels that match the keywords in Appendix \ref{sec:appendix_b}.
However, there are also Dapps that contain buttons that do not match any of our keywords (8\%) or which represent their buttons as images (8\%). 15\% require users to give their consent by ticking a checkbox before being able to interact with them. Finally, 7\% require users to create an account and login via email and password.

Moreover, during our crawl we only visited the landing page of a website or DApp and might have missed any third-party scripts that perform tracking or leak the user's wallet address. This and the fact that we were not able to connect to all the DApps and limited ourselves to a handful of wallets (extensions are well as simulated wallet APIs), highlights that our results should only be considered as a lower bound.

\subsection{Countermeasures}

Privacy-conscious users want to prevent their wallet address to be leaked to third-parties, but also minimize their footprint (i.e., the fact that they have a wallet installed on their browser) for online tracking. 
Initially, wallets would expose the user's unique wallet address to any website without asking the user for prior permission. However, this changed with the release of EIP-$2255$~\cite{eip2255} which requires wallets to ask user's for permission prior to returning any sensitive information to DApps. 
However, the permission system is still flawed. Any third-party that is embedded inside the DApp also has access to the sensitive information once, although the user has granted only permission to the DApp and not the third-parties.

Winter et al.~\cite{winter2021web3} proposed a countermeasure which does not prevent wallet address leakage per-se, but limits its usefulness in linking users across DApps as it generates individual wallet addresses for each DApp a user visits. This follows a similar idea that has been proposed in the past to prevent linking users across websites by using different yet consistent web identities across websites~\cite{TorresJM15}.
Specifically, DApps always interact with a fake proxy wallet address that is derived from the user's real wallet address. All requests that either go through MetaMask or via an JSON-RPC provider are then intercepted and the fake address is swapped with the user's real wallet address such that the DApp is able to perform actions on real data. 

However, this approach has several pitfalls. 
First, the fact that the user's real balance is returned allows DApps and other third-parties to map the fake address to the real address by scanning the blockchain for an address that has the exact same balance. This is trivial because the balance has a high resolution (e.g., $256$-bit resolution in the case of Ethereum) and thus the likelihood that two users having the exact same balance is very low. 
Second, the interception of traffic as well as the swapping between fake and real requires complex management and is prone to errors. For example, transactions are usually not directly mined and most DApps rely on a transaction receipt which includes a transaction hash that allows them to continuously poll the blockchain for the transaction's confirmation status. Hence, the countermeasure also needs to fake transaction hashes otherwise DApps and third-parties might use this information to obtain the users real wallet address. But in fact this might break the usability of many DApps as they sometimes point to other websites such as Etherscan using the transaction hash.
Third, the proposed countermeasure does not hide the existence of a wallet extension from third-parties. Trackers will still be able to detect whether or not a user has a wallet installed on its browser.

As an alternative, users could rely on Ad blockers \cite{disconnect,easylist,whotracksme,duckduckgoBlocklist} to simply block requests from and to third-party tracking scripts. In Section \ref{sec:tracking_results}, we measured  the effectiveness of popular blocklists against the third-parties that we found to access wallet information. The best performing blocklist only managed to block $46$ out of $118$ third-parties (i.e., $39\%$).
Moreover, even with all the blocklists combined, only $51\%$ of the third-parties are blocked. Blocklists do not scale, they can simply be evaded by deploying the same script to a different domain that is not yet blacklisted. For instance, we found that the script which is hosted on \url{wpadmngr.com} (top $1$ in our list of detected third-party scripts) is identical to the scripts hosted on \url{ba0182aa75.com} and \url{6347032d45.com}. Since the two last domains appear to be random, we assume that they might be used by \url{wpadmngr.com} to avoid blocklists.

%% file: sections/related_work.tex
\section{Related Work}


There are several ways to track users online, ranging from classical stateful methods such as third-party cookies \cite{AcarEEJND14,EnglehardtREZMN15,LernerSKR16} to novel stateless methods such as browser fingerprinting \cite{BodaFGI11,NikiforakisKJKPV13,LaperdrixRB16}. 
A number of studies have been conducted over the past years in order to measure the prevalence of third-party cookies and novel browser fingerprinting techniques \cite{Gomez-BoixLB18,VastelLRR18,LaperdrixBBA20,IqbalES21}.
Essentially, any JavaScript API that provides stable yet user-configuration specific information can be leveraged to generate together with other attributes a unique browser fingerprint. This information may range from simple properties such as screen resolution to more advanced techniques such as canvas fingerprinting~\cite{AcarJNDGPP13}. 
For instance, Englehardt et al.~\cite{englehardt2016census} were the first to provide evidence that third-party trackers enhance their browser fingerprinting scripts with information provided by the WebRTC API, Audio API, and Battery Status API. Our work analyses whether trackers are leveraging wallet APIs to enhance their browser fingerprinting scripts to better track users online.

Recently, Senol et al.~\cite{SenolAHB22} discovered that a large number of websites leak the user's email address and password to third-parties. In a similar vein, our work aims to shed light into the inner workings of DApps and wallets to uncover if they might leak a user's wallet address or password to third-parties.

Privacy is not only difficult to achieve on the web, but it is also challenging to achieve when dealing with cryptocurrencies.
Security and privacy concepts are often not well understood by cryptocurrency users. For instance, Krombholz et al.~\cite{KrombholzJGW16} surveyed over $900$ users with respect to their knowledge on security and privacy of Bitcoin. None of the users made a backup of their wallet passphrase on a separate computer. $22\%$ report that they already lost some of their cryptocurrency due to scams or loss of their passphrase. Also, $32\%$ think that Bitcoin is anonymous, despite the fact that transactions can be traced. This is in line with the findings of Mai et al.~\cite{MaiPGWK20} and Voskobojnikov et al.~\cite{VoskobojnikovOH20} where users do not understand the concept of public and private keys or believe that transactions are confidential and cannot be seen by third-parties. These works point out that users might have a misconception of wallets with respect to the privacy that they provide.

As more and more online vendors accept cryptocurrencies as a payment method and an increasing number of decentralized applications begin to emerge, the question around linkability and user privacy becomes indispensable.
A number of previous works have focused on analyzing the linkability of cryptocurrency transactions \cite{MeiklejohnPJLMV16,ChanO17,AzharW20} including their deanonymization via network-layer attacks \cite{BiryukovT19,ApostolakiMV21}. Goldfeder et al.~\cite{GoldfederKRN18} were the first to analyze the intersection between cryptocurrencies and online privacy. The authors find that online trackers are able to collect enough information to link cryptocurrency transactions to online purchases. Béres et al.~\cite{BeresSBQ21} demonstrate how attackers can link different Ethereum addresses to the same user by analyzing meta information such as time of the day and gas price. Even mixers (i.e., services that shuffle transactions in order to break linkability) have been found to be broken \cite{HongKLH18,GhesmatiFW22,abs-2204-02019}. Users often do not understand how to use mixers properly and use, for example, the same wallet address for depositing and retrieving cryptocurrency, thereby making mixing essentially useless. 

Li et al.~\cite{LiCLT0L21} present a denial-of-service attack against blockchain providers, which are frequently used by DApps and wallets to retrieve blockchain information. Blockchain providers often do not impose a gas limit on certain operations and thus malicious users may exploit this fact to make blockchain providers engage in heavy computations.

The work by Winter et al.~\cite{winter2021web3} is the closest to our work. However, their goal is to analyze the security, privacy, and decentralization properties of popular DeFi front ends, while we aim to analyze the privacy implications of wallets. The authors analyzed $78$ handpicked DeFi websites for wallet address leakage and found that $17\%$ of the websites leak the user's wallet address. We found that $59\%$ of the websites leak the user's wallet address. This is because our framework not only analyzes HTTP GET requests but also HTTP POST requests, WebSockets, and cookies. 
Moreover, while Winter et al.~analyzed the websites manually, our work analyzes them automatically. This enables us to perform an automated large-scale study on DApps. Finally, Winter et al.~did not analyze whether wallet extensions also leak the user's wallet address and whether websites make use of wallet information to fingerprint users.

%% file: sections/conclusion.tex
\section{Conclusion}

We present the first systematic study on Web3-based browser fingerprinting and wallet address exfiltration. We built a framework which is capable of detecting JavaScript calls on wallet APIs as well as intercept and search HTTP requests, WebSockets and cookies for leaked wallet addresses.
Our framework integrates a wallet simulator which imitates different wallet extensions by injecting wallet-specific properties into the website's DOM, and developed an automator which automatically sets up MetaMask and connects it to DApps. 
Using our framework we analyzed the top 100K websites and found evidence of 1,325 websites checking the presence of wallet extensions installed within the user's browser. We analyzed 1,572 DApps and found that 211 of them leak the user's wallet address to third-parties. Moreover, we analyzed 100 popular wallets and found that 13 of them deliberately leak the user's wallet address to third-parties.
We evaluated countermeasures such as Ad blockers and found that they are not completely effective in blocking all the third-party scripts and leaks detected by our framework. We conclude that wallets pose a serious threat to user's privacy and that new solutions need to be developed that allow users to interact with DApps in a secure and privacy-preserving way.

%% file: sections/appendix.tex
\appendix
\section*{Appendix}

\addcontentsline{toc}{section}{Appendices}
\renewcommand{\thesubsection}{\Alph{subsection}}

\subsection{Browser Fingerprinting Categories}
\label{sec:appendix_a}

\tablename{}~\ref{tab:fingerprinting_categories} lists all the JavaScript APIs that our framework uses to detect browser fingerprinting, including the category that we assigned to each API. For example, \texttt{*userAgent*} means that any JavaScript API call that includes the string ``userAgent'' will be collected and assigned to the category browser, and the category browser is not considered as an explicit browser fingerprinting category. 

\begin{table}
    \centering
    \begin{adjustbox}{width=\columnwidth,center}
    \begin{tabular}{l r c}
    \toprule
    \textbf{JavaScript API Call} & \textbf{Category} & \textbf{Explicit}\\
    \midrule
    
    \texttt{window.ethereum} & Wallet & \xmark \\
    \texttt{window.cardano} & Wallet & \xmark \\
    \texttt{window.solana} & Wallet & \xmark \\
    \texttt{window.BinanceChain} & Wallet & \xmark \\
    
    \texttt{RTCPeerConnection*} & RTC & \cmark \\
    \texttt{RTCPeerConnectionIceEvent*} & RTC & \cmark \\
    
    \texttt{WebGLRenderingContext*} & WebGL & \cmark \\
    
    \texttt{HTMLCanvasElement*} & Canvas & \cmark \\
    \texttt{CanvasRenderingContext2D*} & Canvas & \cmark \\
    
    \texttt{*Storage*} & Storage & \xmark \\
    \texttt{*indexedDB*} & Storage & \xmark \\
    
    \texttt{Screen*} & ScreenSize & \xmark \\
    \texttt{*screen*} & ScreenSize & \xmark \\
    
    \texttt{*cookie*} & Cookies & \xmark \\
    
    \texttt{Date*} & DateTime & \xmark \\
    \texttt{*DateTimeFormat*} & DateTime & \xmark \\
    
    \texttt{*getBattery*} & Battery & \cmark \\
    
    \texttt{*Height*} & WindowSize & \xmark \\
    \texttt{*Width*} & WindowSize & \xmark \\
    \texttt{BarProp*} & WindowSize & \xmark \\
    
    \texttt{*connection*} & Connection & \xmark \\
    \texttt{*onLine*} & Connection & \xmark \\
    
    \texttt{*devicePixelRatio*} & ScreenResolution & \xmark \\
    
    \texttt{*window.name*} & WindowLocation & \xmark \\
    
    \texttt{*plugins*} & Plugins & \cmark \\
    \texttt{*mimeType*} & Plugins & \cmark \\
    \texttt{*canPlayType*} & Plugins & \cmark \\
    
    \texttt{*vendor*} & Browser & \xmark \\
    \texttt{*product*} & Browser & \xmark \\
    \texttt{*platform*} & Browser & \xmark \\
    \texttt{*app*} & Browser & \xmark \\
    \texttt{*userAgent*} & Browser & \xmark \\
    
    \texttt{*language*} & Language & \xmark \\
    
    \texttt{DeviceOrientationEvent*} & Device & \cmark \\
    \texttt{DeviceMotionEvent*} & Device & \cmark \\
    \texttt{*maxTouchPoints*} & Device & \cmark \\
    \texttt{*hardwareConcurrency*} & Device & \cmark \\
    \texttt{*deviceMemory*} & Device & \cmark \\
    \texttt{*memory*} & Device & \cmark \\
    
    \texttt{AudioBuffer*} & Audio & \cmark \\
    \texttt{OfflineAudioContext*} & Audio & \cmark \\
    
    \texttt{*requestMediaKeySystemAccess*} & Media & \xmark \\
    \texttt{*mediaDevices*} & Media & \xmark \\
    \texttt{*enumerateDevice*} & Media & \xmark \\
    \texttt{*mediaCapabilities*} & Media & \xmark \\
    
    \texttt{Navigator*} & Navigator & \xmark \\
    
    \texttt{Performance*} & Performance & \xmark \\
    
    \texttt{speechSynthesis*} & SpeechSynthesis & \cmark \\

    \bottomrule
    \end{tabular}
    \end{adjustbox}
    \caption{Browser fingerprinting related JavaScript API calls and assigned category.}
    \label{tab:fingerprinting_categories}
\end{table}

\subsection{List of Keywords Used by Automator}
\label{sec:appendix_b}

\tablename{}~\ref{tab:automator_keywords} lists all the keywords that our automator scans for within a website's HTML to find a ``Connect'' and ``MetaMask'' button.

\begin{table}[h]
    \centering
    \begin{tabular}{l p{6cm}}
    \toprule
    Connect & \emph{``Connect to MetaMask''}, \emph{`` Connect Wallet ''}, \emph{``Connect Wallet''}, \emph{``Connect wallet''}, \emph{``connect wallet''}, \emph{``Connect to a wallet''}, \emph{``Connect to wallet''}, \emph{``Connect your wallet''}, \emph{``Sign In''}, \emph{``Connect''}, \emph{``CONNECT WALLET''}, \emph{``CONNECT''}, \emph{``SIGN IN''}, \emph{``WALLET''}, \emph{``SIGN''}, \emph{``sign''}, \emph{``SIGNIN''}, \emph{``Sign Up''}, \emph{``Connect Your Wallet''}, \emph{``Wallet''}, \emph{``Connect a Wallet''}, \emph{``Connect a wallet''}, \emph{``Sign in''}, \emph{``sign in''}, \emph{``connect''}, \emph{``Log in via web3 wallet''}, \emph{``wallet''}, \emph{``account''}, \emph{``Account''} \\
    \midrule
    MetaMask & \emph{``MetaMask''}, \emph{``MetaMask ''}, \emph{``metamask''}, \emph{``Connect Metamask''}, \emph{``Connect MetaMask''}, \emph{``Metamask''}, \emph{``Connect to MetaMask''}, \emph{``browser wallet''}, \emph{``Browser Wallet''}, \emph{``Browser wallet''}, \emph{``Metamask \& Web3''} \\
    \bottomrule
    \end{tabular}
    \caption{Keywords used by the automator to identify connect and MetaMask buttons on DApp websites.}
    \label{tab:automator_keywords}
\end{table}

\subsection{Breakdown of Connection Failures}
\label{sec:appendix_c}

\tableautorefname{} \ref{tab:connection_failures} provides a breakdown over the reasons that resulted in our automator in failing to automatically connect to the DeFi related DApps from our DappRadar.com dataset.

\begin{table}[h]
    \centering
    \footnotesize
    \begin{tabular}{l r r}
    \toprule
\textbf{Connection Failure} & \textbf{\#} & \textbf{(\%)} \\
\midrule
Not a DApp & 32 & (24\%) \\
Button text not detectable & 24 & (18\%) \\
Consent required & 20 & (15\%) \\
Website down & 19 & (14\%) \\
Different button label & 11 & (8\%) \\
Button is an image & 10 & (8\%) \\
Login required & 9 & (7\%) \\
MetaMask not supported & 4 & (3\%) \\ 
Requires blockchain network selection & 3 & (2\%) \\
Captcha & 1 & (1\%) \\
\midrule
\textbf{Total} & \textbf{133} & \textbf{100\%} \\
\bottomrule
    \end{tabular}
    \caption{Breakdown of reasons of connection failures by our automator. The breakdown is based on the DeFi subset of our DappRadar.com dataset.}
    \label{tab:connection_failures}
\end{table}